\documentclass{emulateapj}
\usepackage{multirow}
\usepackage{natbib}

\usepackage{hyperref}
\usepackage[UKenglish]{babel}

\begin{document}

\title{Cooling, AGN Feedback and Star Formation in Simulated Cool-Core Galaxy Clusters}

\author{Yuan Li\altaffilmark{1}, Greg L. Bryan\altaffilmark{2}, Mateusz Ruszkowski\altaffilmark{1}, G. Mark Voit\altaffilmark{3}, Brian W. O'Shea\altaffilmark{3,4}, Megan Donahue\altaffilmark{3}}

\altaffiltext{1}{Department of Astronomy, University of Michigan, 1085 S University Ave, Ann Arbor, MI 48109; email: yuanlium@umich.edu}
\altaffiltext{2}{Department of Astronomy, Columbia University, Pupin Physics Laboratories, New York, NY 10027}
\altaffiltext{3}{Department of Physics and Astronomy, Michigan State University, East Lansing, MI 48824}
\altaffiltext{4}{Lyman Briggs College and National Superconducting Cyclotron Laboratory, Michigan State University, East Lansing, MI 48824}

\begin{abstract}
Numerical simulations of active galactic nuclei (AGN) feedback in cool-core galaxy clusters have successfully avoided classical cooling flows, but often produce too much cold gas. We perform adaptive mesh simulations that include momentum-driven AGN feedback, self-gravity, star formation and stellar feedback, focusing on the interplay between cooling, AGN heating and star formation in an isolated cool-core cluster. Cold clumps triggered by AGN jets and turbulence form filamentary structures tens of kpc long. This cold gas feeds both star formation and the supermassive black hole (SMBH), triggering an AGN outburst that increases the entropy of the ICM and reduces its cooling rate. Within 1-2 Gyr, star formation completely consumes the cold gas, leading to a brief shutoff of the AGN. The ICM quickly cools and redevelops multiphase gas, followed by another cycle of star formation/AGN outburst. Within 6.5 Gyr, we observe three such cycles. There is good agreement between our simulated cluster and the observations of cool-core clusters. ICM cooling is dynamically balanced by AGN heating, and a cool-core appearance is preserved. The minimum cooling time to free-fall time ratio typically varies between a few and $\gtrsim 20$. The star formation rate (SFR) covers a wide range, from 0 to a few hundred $\rm M_{\odot}\, yr^{-1}$, with an average of $\sim 40 \,\rm M_{\odot}\, yr^{-1}$. The instantaneous SMBH accretion rate shows large variations on short timescales, but the average value correlates well with the SFR. Simulations without stellar feedback or self-gravity produce qualitatively similar results, but a lower SMBH feedback efficiency (0.1\% compared to 1\%) results in too many stars.

\end{abstract}


\section{Introduction}
X-ray observations of the intracluster medium (ICM) indicate that in the cores of some galaxy clusters, the gas with high density and low entropy has a cooling time much shorter than the Hubble time \citep[e.g.][]{Fabian1977}. These clusters are often called ``cool-core clusters.'' In the absence of heating, cold gas is expected to condense out of the ICM in the center of cool-core clusters, driving an inflow (the so-called ``cooling flow'') at rates of 100s-1000 $\rm M_{\odot} yr^{-1}$ \citep[see review by][]{Fabian1994}. However, more recent X-ray observations made by Chandra and XMM-Newton have revealed a dearth of cooler X-ray gas below 1/3 of the virial temperature \citep[e.g.][]{Peterson2003, Peterson2006}, and the observed star formation rate is usually much lower than the classical cooling rate \citep[e.g.][]{McNamara1989, ODea2008}. The lack of observational support for the existence of a classical cooling flow is referred to as the ``cooling flow problem.'' The solution to this problem usually involves some heating mechanism that can balance radiative cooling \citep[e.g.,][]{Ruszkowski2002, Ruszkowski2004, Kim2005, Heinz2006, Brighenti2006, GO2008, Conroy}. 

It is widely accepted that the feedback from AGNs is the major heating source in cool-core clusters \citep{McNamara2007}. X-ray cavities created by AGN outbursts are often observed in nearby cool-core clusters \citep[e.g.,][]{Fabian2006, Wise2007, Baldi2009, Blanton2011}, and the energy associated with the cavities are usually enough to offset cooling \citep{Dunn2006, Rafferty2006}. There is mounting observational evidence linking the AGN activities, the cooling properties of the ICM, and the multiphase gas: clusters with low entropy X-ray gas (i.e., with a short cooling time) in the cores almost always harbor line-emitting multiphase gas and radio-loud AGN, while the clusters with hotter cores usually do not \citep{Best2007, Cavagnolo2008, Mittal2009, Birzan2012}.

Numerical simulations have shown that AGN feedback in the form of momentum-driven jets can successfully suppress cooling in both cosmological simulations \citep[e.g.][]{Dubois2010, Martizzi2012} and isolated systems \citep[e.g.][]{Cattaneo2007, Gaspari2011}. Recent simulations with ``cold feedback'' in which the AGN is powered by the accretion of cold gas \citep{Pizzolato2005, Gaspari2013}  have not only successfully achieved long-term global thermal equilibrium with a range of feedback efficiencies, but also produced multiphase filamentary structures (reaching radii of up to tens of kpc) that morphologically resemble the observations of nearby cool-core clusters \citep[e.g.][hereafter LB14a]{Gaspari2012, PIII}. However, multiphase gas usually only exists at early times of the simulations (up to 1-2 Gyr), and due to the conservation of angular momentum, much of the gas that has initially cooled forms a massive cold disk of $\sim10^{11}~\mathrm{M}_{\odot}$ that persists for many Gyr through the end of the simulation \citep[][hereafter LB14b] {PII}. Observations of optical and IR line emission shows great diversity in the properties of multiphase gas in cool-core clusters: some systems host extended filamentary cold gas like the Perseus Cluster \citep{Fabian2003, Fabian2006}; some only have detections in the nuclei; some do not seem to harbor any cold gas \citep{Edwards2009, McDonald10, McDonald11}. The amount of cold gas detected is also typically smaller than $10^{11}~\mathrm{M}_{\odot}$ \citep{Edge2001, Salome2003}. This suggests that the phase with a long-lived massive cold disk is unrealistic and additional physics is needed in the simulations.

Star formation is an important process that is often overlooked in simulations focused on suppressing cooling flows with AGN feedback. This is mainly because stellar feedback is energetically insufficient to balance cooling except under extreme assumptions \citep{Bregman1989, Skory}. Cosmological simulations \citep[e.g.,][]{Sijacki2006, Dubois2010} show that star formation is suppressed but still occurs when AGN feedback is included. Numerous observations show that a significant fraction of brightest cluster galaxies (BCGs) in cool-core clusters are forming stars \citep{Johnstone1987, Cardiel1995, Cardiel1998, Crawford1999, Hicks2005, Edwards2007}. There is a general consensus that a ``reduced cooling flow'' is present in most cool-core clusters: some of the hot ICM cools into (in some systems filamentary) cold structures; some of this cold gas forms stars, at a rate that is on average an order of magnitude lower than the classical rate, but with a large variation, ranging from below detection limit up to the classical rate of hundreds of solar masses per year \citep{Hicks2010, McDonald11b, Hoffer2012, Phoenix}. Star formation is also largely responsible for ionizing the line-emitting filaments \citep{ODea2010, McDonald12}. However, exactly how star formation in the BCGs is connected to the cooling of the ICM is not well understood, and there is a lack of detailed modeling of star formation in these largest galaxies.

In this paper, we carry out three-dimensional adaptive mesh refinement (AMR) simulations to study the interplay between ICM cooling, star formation and AGN feedback in an idealized cool-core galaxy cluster. The model is based on our previous work that focused on the balance between cooling and momentum-driven AGN feedback (LB14a \& LB14b). Besides cooling and AGN feedback, we also include the self-gravity of the gas, star formation, and stellar feedback in the standard run presented here. The goal of this work is to gain a better understanding of the following issues: (1) How does AGN suppress ICM cooling and mediate star formation? (2) What is the impact of star formation and stellar feedback on the AGN activities? (3) Given the self-regulating nature of AGN feedback, can we still put constraints on the feedback efficiency? (4) What is the physical explanation for the observed relation between the ICM properties, AGN activities, the multiphase gas, and star formation? 

We structure the paper as follows: in Section~\ref{sec:method}, we describe the methodology of this work including the refinement criteria, the setup of the initial conditions, and the modeling of AGN feedback, star formation, and stellar feedback; in Section~\ref{sec:results}, we describe the results from the standard simulation, analyze the intertwined relationship between cooling, AGN and star formation, discuss our parameter studies and other test runs we have preformed to understand the role of individual pieces of physics included in the simulations; in Section~\ref{sec:discussion} we compare our results to the observations and other related simulation work, and discuss the potential effect of other physical processes such as thermal conduction. We conclude this work in Section~\ref{sec:conclusion}.

\section{Methodology}
\label{sec:method}

Our three-dimensional simulations are performed using the AMR code Enzo \citep{Enzo}, with the ZEUS hydrodynamic method \citep{Zeus}. All of the simulations discussed in this paper have $64^3$ root grids ($N_{root}$) and 10 maximum refinement level ($l_{max}$) in a box of $L=16$ Mpc. This gives the smallest cell size of $\Delta x_{\rm min} = L / (N_{\rm root} 2^{l_{\rm max}}) \approx 244$ pc. The resolution is lower than LB14a and LB14b (with $\Delta x_{\rm min} \sim 15$ and $\sim 60$ pc), but as is shown in LB14a, the general results are converged as long as $\Delta x_{\rm min} \lesssim 500$ pc. The refinement strategy is kept the same as our previous work and is briefly outlined here \citep[a detailed description can be found in][]{P1}. A cell is refined if any of the following three criteria are met: (1) the gas mass in the cell exceeds 0.2 times the mean mass of the gas in one cell of the root grid; (2) the ratio of the cooling time to the sound-crossing time is smaller than some limit $\beta$, (we use $\beta=6$ here, a somewhat arbitrary value larger than 1, to better resolve cooling); (3) the size of the cell is larger than 1/4 of the Jeans length (the local Jeans length is always resolved by at least 4 cells in each spatial dimension, to prevent numerically-induced fragmentation of the gas) \citep{Truelove}.

The important physical processes included in the simulations are radiative cooling, momentum-driven AGN feedback and star formation (with feedback). The self-gravity of the gas is also taken into account. The radiative cooling curve is the same as that used in \citet{Tasker2006}, with a temperature floor of 300 K. Throughout the paper, we refer to the diffuse ICM with temperatures above $10^7$ K as the ``hot'' gas, and the condensed gas with temperatures below $10^5$ K as the ``cold'' gas. The simulation data is analyzed using the yt package \citep{yt}. 

We describe the cluster setup in Section ~\ref{sec:method_1}. The modeling of AGN jets and star formation are described in Section ~\ref{sec:method_2} and Section ~\ref{sec:method_3}, respectively. %

\subsection{Cluster Setup}\label{sec:method_1}

The initial setup is very similar to \citet{P1}, LB14a and LB14b. Our idealized isolated cool-core galaxy cluster is constructed based on the observations of the Perseus Cluster. Following \citet{Mathews}, the electron density profile is initially set to be:
\begin{equation}
n_e(r) = \Bigg(\frac{0.0192}{1 + \left(\frac{r}{18}\right)^3} + \frac{0.046}{\left[1 + \left(\frac{r}{57}\right)^2\right]^{1.8}}
+ \frac{0.0048}{\left[1 + \left(\frac{r}{200}\right)^2\right]^{1.1}}\Bigg)\,\rm{cm}^{-3},
\end{equation}
where $r$ is the distance to the cluster center in kpc. The power-law index of the last term is slightly steepened so that the density profile at large radii is more consistent with cosmological simulations as well as the observations of the outskirts of Perseus \citep{Urban2014}. Because our focus is on the cluster core, and both cooling and dynamical timescales are very long outside the core, the exact value of the index is unimportant. 

The initial temperature profile within $r < 300$ kpc is taken from observations \citep{Churazov}:
\begin{equation}
T = 7 \; \frac{1 + (r/71)^3}{2.3 + (r/71)^3} \; \rm{keV}\;,
\end{equation}
while at larger radii, we adopt the universal temperature profile found by \citet{Universal_T}, normalizing it to match the observations at $r = 300$ kpc:
\begin{equation}
T = 9.18 \Big(1+\frac{3\;r}{2\;r_{\rm vir}}\Big)^{-1.6}\; \rm{keV}\;,
\end{equation}
where $r_{\rm vir} = 2.440$ Mpc is the virial radius of the cluster. 

The brightest cluster galaxy (BCG) is treated as a fixed potential \citep{Mathews} with the stellar acceleration: 
\begin{equation}
g_*(r) = 
 \left[ \left( {r^{0.5975} \over 3.206 \times 10^{-7}}\right)^{0.9}
+ \left( {r^{1.849} \over 1.861\times 10^{-6}}\right)^{0.9} 
\right]^{-1/0.9}\;.
\end{equation}

The SMBH in the center of the cluster is treated as a point mass of $M_{\rm SMBH} = 3.4 \times 10^8$ M$_{\odot}$ \citep{BHmass}.

We assume an ideal gas law for the ICM with $\gamma = 5/3$, and that the ICM is initially in hydrostatic equilibrium with the gravitational potential which includes the contribution from the dark matter, the BCG, the SMBH, and the gas itself. This allows us to fit an NFW profile \citep{NFW} to the dark matter halo:
\begin{equation}
\rho(r)=\frac{\rho_0}{\frac{r}{R_s}(1+\frac{r}{R_s})^2}\;,
\end{equation}
where $\rho_0$ is found to be $8.42\times 10^{-26}\, \mathrm{g}\, \mathrm{cm}^{-3}$, and the scale radius $R_s = 351.7$ kpc.

We do not include any initial rotation or perturbation in the gas.

\subsection{Jet Modeling}\label{sec:method_2}

The SMBH accretion and jet modeling is the same as in LB14a \& b. Here we only repeat the key aspects.

We calculate the accretion rate $\dot{M}_{\rm SMBH}$ at each time step by dividing the total amount of cold gas in the close vicinity ($r<500$ pc) of the SMBH by a typical accretion time (5 Myr). Then a fraction of the cold gas mass within the accretion region is removed in proportion to the cell mass. 

The AGN jets in our simulations are modeled as bipolar outflows along the $z$-axis from two parallel jet launching planes \citep{Omma2004}. The jet launching planes are parallel to the x-y plane, at a distance of 2 cell lengths from the SMBH located at the center of the simulation domain. Within the jet launching planes, the mass added to the cells at each time step $\Delta t$ follows $\Delta m \propto e^{-r^2/2r^2_{\rm jet}}$, where r is the distance to the 
$z$-axis and $r_{jet} = 1.5 \Delta x_{\rm min}$. Assuming only a small fraction of $\dot{M}_{\rm SMBH}$ is added onto the SMBH for radio mode AGN, the outflow rate is approximately equal to the accretion rate $\dot{M}_{\rm SMBH}$. Therefore $\Delta m$ can be normalized by setting $\int \Delta m =\dot{M}_{\rm SMBH} \Delta t$.  Our simulations are far from being able to resolve the accretion physics. It is even possible that only a fraction of this accreted gas actually reaches the accretion disk, and the rest is pushed out by the jets. Strictly speaking, $\dot{M}_{\rm SMBH}$ is the rate at which gas is processed by the SMBH, but for simplicity, we refer to it as the SMBH accretion rate throughout the paper.

The jet power is expressed as
\begin{equation}\label{eq:Edot}
\dot{E} = \epsilon  \dot{M}_{\rm SMBH} c^2 \; ,
\end{equation}
where $\epsilon$ is the feedback efficiency. In our standard run, we use $\epsilon=1\%$. We will discuss a test run with a lower feedback efficiency, $\epsilon=0.1\%$, in Section~\ref{sec:othersims_2d}. In all simulations performed here, we assume the jet is $50\%$ thermalized. Thus the kinetic power of the jet is $\dot{E}_{\rm kinetic}=\dot{E}-\dot{E}_{\rm thermal}=f_{\rm kinetic}\dot{E} = 0.5\dot{E}$. As is shown in LB14a, varying $f_{\rm kinetic}$ from 0.1 to 1 does not significantly affect the results. To avoid the energy loss through low-density channels \citep[e.g.,][]{Reynolds06}, we enforce a small angle ($\theta=0.15$) precession on the jet with a period of 10 Myr. 

\subsection{Star Formation and Stellar Feedback}\label{sec:method_3}

The star formation is modeled following \citet{CenOstriker}. A star particle is created in a cell when the following set of criteria is met: 1) the gas density exceeds a certain value ($1.67\times 10^{-24} \, \mathrm{g} \, \mathrm{cm}^{-3}$ is used in our simulations), 2) the gas mass exceeds the local Jeans mass, 3) the flow is convergent, and 4) the cooling time of the gas is shorter than the dynamical time of the gas in the cell ($t_{\rm dyn}=(3\pi/32G\rho_{cell})^{1/2}$). When all of the criteria are met, a fraction of the gas within the cell is converted into a star particle:
\begin{equation}\label{eq:SFR}
m_*=\epsilon_{\rm SF}\frac{\Delta t}{t_{\rm dyn}}\rho_{\rm gas}\Delta x^3 \; ,
\end{equation}
where $\epsilon_{\rm SF}$ is the star formation efficiency and is set to be 0.02 in our simulations. If $m_*$ is above the minimum mass for star particles, $m_{*,min}$, then the star particle is created. When $m_* < m_{*,min}$, a star particle is created with a probability of $m_* / m_{*,min}$, and its mass would be $m_{*,min}$ or $80\%$ of the gas mass in the cell, whichever is smaller. This minimum mass is set to prevent the formation of too many star particles which would slow down the computation, and our standard run uses $ m_{*,min} = 10^6 \rm M_{\odot}$. 

The star particle is created over a period $\tau$ following:
\begin{equation}
m_*(t)=m_*\int_0^{t_0} \frac{t-t_0}{\tau^2} \exp(-\frac{t-t_0}{\tau})dt\; ,
\end{equation}
where $\tau $ is the maximum of the dynamical time $t_{\rm dyn}$ and 10 Myr.

We also include the stellar feedback from mostly Type II SNe by injecting mass and energy back into the gas over a dynamical time. $25\%$ of the mass of the created stars is returned to the gas phase as stellar winds and SNe ejecta. The feedback in the form of thermal energy from each star particle is $\epsilon_{\rm SNe}m_*c^2$, where $\epsilon_{\rm SNe}$ is set to be $10^{-5}$. Both mass and energy feedback are injected locally to the cell that contains the star particle. We discuss how stellar feedback and the choice of star formation parameters affect the results in Section~\ref{sec:results_3}.

\section{Results}\label{sec:results}
In this section we present the main results of the simulations. Section~\ref{sec:results_1} describes the cluster evolution in the standard run, and in Section~\ref{sec:results_2} we examine the intricate relationship between cooling, star formation, and AGN feedback.  In Section~\ref{sec:results_3} we discuss other simulations that we have performed to study the role each physical process plays and the impact of changing some of the important simulation parameters.

\subsection{Cluster Evolution}\label{sec:results_1}
The cluster evolution in the first few hundred Myr is similar to that in LB14a \& b: about 300 Myr ($\sim$ the initial minimum cooling time of the gas) after we start the simulation, runaway cooling happens first in the very center of the cluster. The cold gas is accreted onto the SMBH and turns on the AGN jets; the jets then perturb the ICM in a non-linear fashion, dragging out low entropy gas with short cooling time to larger radii, causing it to cool into clumps along the jet path. A detailed analysis of this cooling process can be found in LB14b. These cold, dense clumps have a lower pressure than the surrounding ICM, which drives more hot gas to cool onto them, but as the clumps move through the hot ICM they also suffer from shock heating, ram pressure stripping and Kelvin-Helmholtz instabilities, which can reduce their masses or even destroy them completely. The fate of individual clumps is uncertain (as they are numerical difficult to model accurately), but at the beginning of the clump formation stage the growth generally overwhelms destruction.

\begin{figure*}
\begin{center}
\includegraphics[scale=.34]{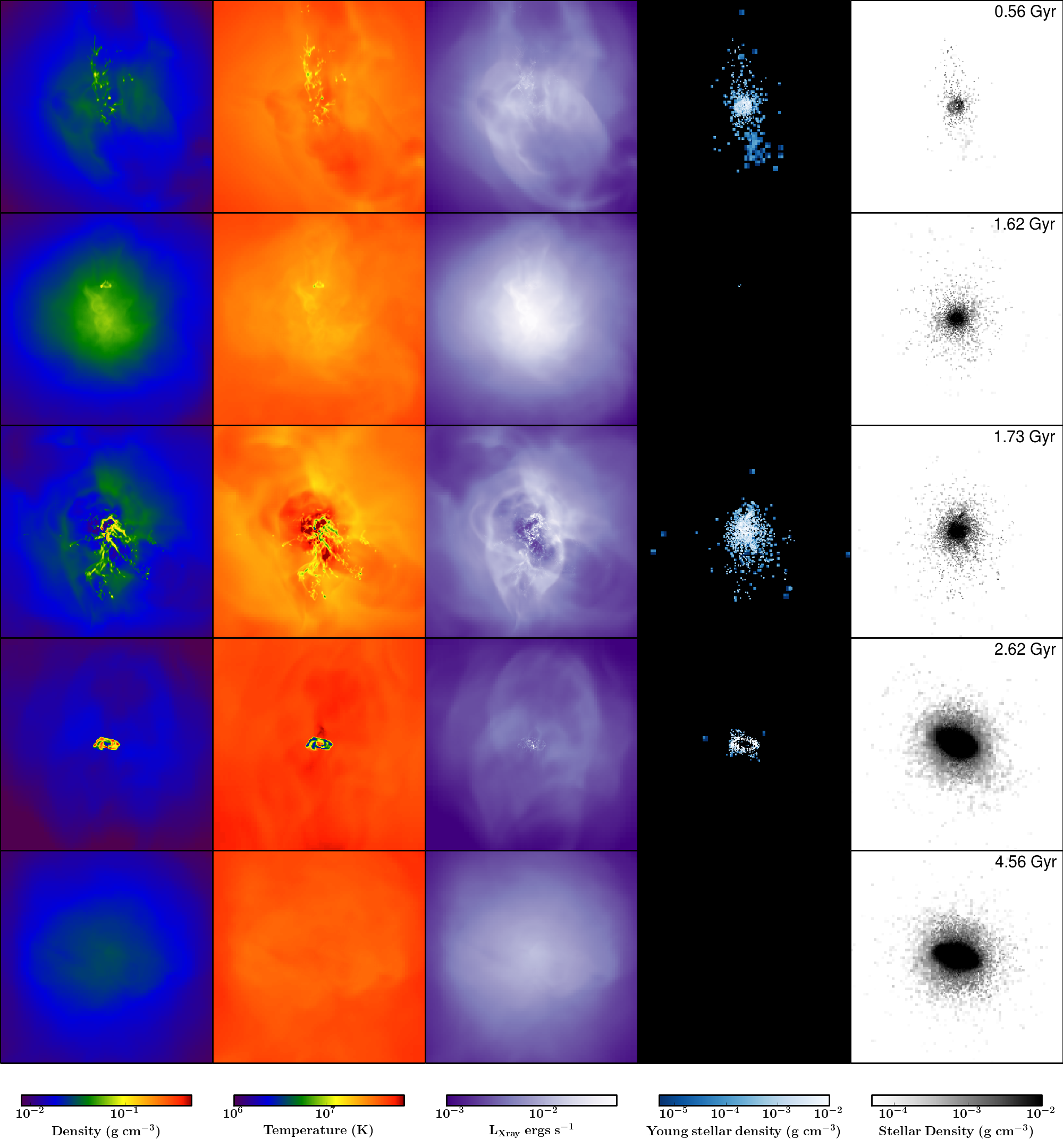}
\caption{The projected gas density, temperature (weighted by density), X-ray luminosity, young stellar density and stellar density in the central $80\times80\times80$ kpc$^3$ box at $t=0.56$, 1.62, 1.73, 2.62 and 4.56 Gyr. The projection is along the $x$-axis and the AGN jets are along the $z$-axis (vertical). The young stellar density only includes star particles less than 200 Myr old, which can be observed in the UV. The stellar density only shows the stars formed in the simulation but not the stars that are already in the BCG at the beginning of the simulation (treated only as a fixed potential). An animation of the temperature evolution can be viewed here: \url{https://vimeo.com/115825854}.
\label{fig:multi}} 
\end{center}
\end{figure*}

The cold clumps fall to the center of the cluster roughly within a dynamical time. Some of the cold gas is accreted onto the SMBH, boosting its power, which increases the entropy of the ICM, halting clump formation. Some of the cold clumps also form stars, and stellar feedback heats up the cold gas locally, reducing its mass. The total amount of cold gas in the first 1.5 Gyr peaks at only $7\times10^{9} \rm~M_{\odot}$ and does not form a rotationally supported disk (the first row of Figure~\ref{fig:multi}). At t $\sim 1.5$ Gyr, all the cold gas has vanished and the AGN shuts off. The cluster enters a state that is very similar to the initial condition except that the gas is now turbulent.

The second cycle is similar to the first one. At the end of the first cycle, the shortest cooling time of the ICM is only a few hundred Myr. With the AGN heating off, the ICM quickly cools again. Due to the turbulent motion of the gas, runaway cooling first happens $\sim 15$ kpc from the cluster center (the second row of Figure~\ref{fig:multi}) from a blob of low entropy gas at t~$= 1.62$ Gyr. The cooled gas shortly gets accreted onto the SMBH, turning the AGN back on. The AGN jets trigger more hot gas to cool into filamentary structures, resulting in a burst of star formation and AGN feedback (the third row of Figure~\ref{fig:multi}). The total amount of cold gas peaks at $5 \times 10^{10} \rm~M_{\odot}$, about one order of magnitude higher than the previous cycle. Cooling out of the ICM at a few tens of kpc, the cold gas carries a non-zero total angular momentum. Due to the larger amount of cold gas in this cycle (and higher total angular momentum), SMBH accretion and star formation cannot consume all of it quickly enough before the gas settles to a rotating disk in the center of the cluster at t$\sim 2.5$ Gyr. The cluster enters a stage that appears very similar to the last stage of the simulations with AGN feedback only (LB14a): the AGN outburst has elevated the entropy of the ICM in the cluster core; all the gas with short cooling time ($\sim$ a few hundred Myr) has already cooled, and thus clump formation has stopped; the ICM cools directly onto the rotationally supported cold disk; the SMBH continues to accrete from the inner part of the cold disk, keeping the AGN jets on, which balances the ICM cooling (the fourth row of Figure~\ref{fig:multi}). In the simulations with AGN only, due to the conservation of angular momentum, a disk of $\sim 10^{11} \rm~M_{\odot}$ lasts through the end of the simulation without showing signs of shrinking (LB14a). In the standard run shown here, however, star formation gradually consumes the cold disk. By $t=3$ Gyr, the mass of the disk has to be reduced to only $\sim10^{10} \, \rm M_{\odot}$. At $t \sim 4.3$ Gyr, the disk is completely gone (the last row of Figure~\ref{fig:multi}). This marks the end of the second cycle.

The third cycle begins at $t=4.8$ Gyr and ends around $t=6$ Gyr. At the end of our simulation at roughly $t\sim 6.5$ Gyr, there is no cold gas, and both AGN and star formation are off, but the ICM is cooling and ready to enter yet another cycle. 

Figure~\ref{fig:multi} also shows that the star forming regions (the fourth column) are spatially correlated with the cold dense gas (first two columns). A comparison between the morphology of the cold gas in our previous higher resolution simulations \footnote{The cold clumps are better resolved with higher resolution and the bipolar cold structure along the jet path is more prominent at the beginning of the AGN outburst. Because we are not focused on the cold structure in this work, a lower resolution is chosen to reduce computational cost.} (LB14a\&b) and the high quality UV images of BCGs in CLASH \citep{Postman2012} is presented in Donahue et al. 2015 (submitted to ApJ). Interestingly, the cycle we see in our simulation, in which cold, filamentary star-forming gas settles into a central disk and then disappears, also echoes the hypothesized dust-cloud evolution sequence in \citet{Lauer2005} based on Hubble observations of 77 early-type galaxies.

\subsection{The Interplay between Cooling, AGN Feedback and Star Formation}\label{sec:results_2}

\begin{figure}[h]
\begin{center}
\includegraphics[scale=.59,trim=0.5cm 0.6cm 0cm 0cm, clip=true]{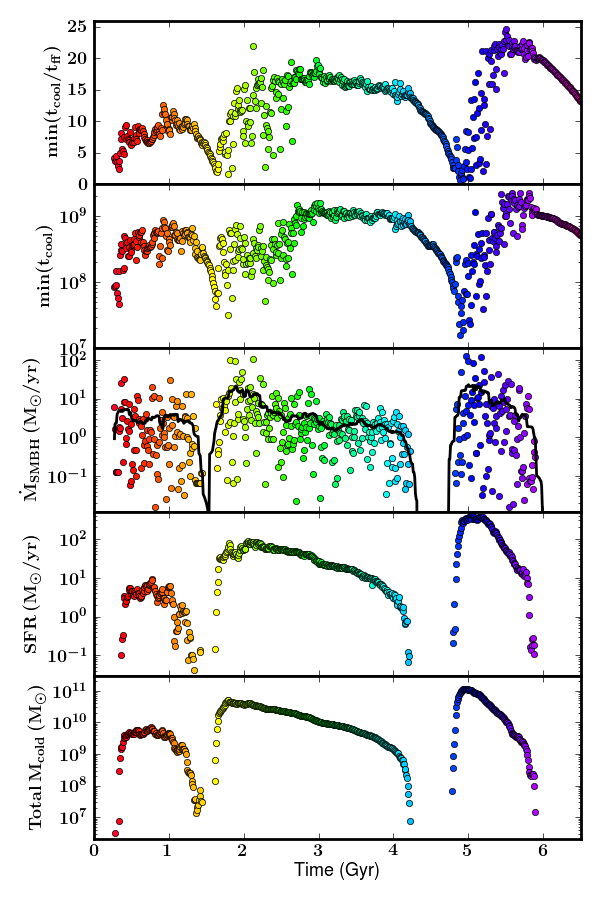}
\caption{The time evolution of the minimum $t_{\rm cool}/t_{\rm ff}$ ratio, the minimum cooling time $t_{\rm cool}$, the SMBH accretion rate $\rm \dot{M}_{\rm SMBH}$, the SFR, and the total amoun{\normalsize {\large {\Large }}}t of cold gas in the system. Data is sampled every 10 Myr and the color indicates time (from red to purple). The black line in the third panel is the $\rm \dot{M}_{\rm SMBH}$ averaged with a 200 Myr moving window. 
\label{fig:5in1}}
\end{center}
\end{figure}

Both AGN and star formation are fueled by the cold gas that cools out of the hot ICM. AGN feedback heats the ICM and regulates its cooling rate, while star formation consumes the cold gas and thus affects the AGN duty cycle. In this section, we study the intricate relationship between the ICM cooling, AGN feedback and star formation.
There are many ways of measuring the cooling properties of the hot ICM in the core of the cluster, such as temperature, central entropy, and classical cooling rate \citep[which is often defined as the cooling rate within the cooling radius where the cooling time is a few Gyr; e.g.,][]{Mittal2009}. It is also found that precipitation in cool-core clusters is tightly linked to the minimum ratio of the cooling time over the free-fall time $t_{\rm cool}/t_{\rm ff}$ \citep[e.g.,][]{MarkMegan2014}. Since central entropy, classical cooling rate, and the minimum ratio convey very similar information (low central entropy translates to high classical cooling rate and low minimum ratio), we mainly use the minimum ratio to characterize the cooling properties of the ICM and we show its evolution with time in Figure~\ref{fig:5in1}, along with the minimum cooling time $t_{\rm cool}$, SMBH accretion rate, SFR and the total amount of cold gas. Note that we use the shell averaged $t_{\rm cool}$ as is done in observations, instead of the actual value measured in each cell.

The minimum $t_{\rm cool}/t_{\rm ff}$ ratio approaches its local minimum at the beginning of every cycle. When AGN feedback is just turned on, this ratio still decreases for a very brief period of time. It takes roughly one dynamical time (a few tens of Myr at $r\sim 10-30$ kpc) for most of the cold clumps to reach the SMBH and trigger the AGN outburst. The instantaneous SMBH accretion rate shows very large variations, but when we take its average with a moving window of 200 Myr, the trend is clear (the black line in the third panel of Figure ~\ref{fig:5in1}). The minimum ratio starts to increase right before the AGN outflow rate (also equal to the SMBH accretion rate in our model) reaches its peak. In every cycle, within 1 Gyr of the time that the AGN is turned on, all the gas with short $t_{\rm cool}$ of a few hundred Myr has either cooled or been heated to higher temperatures by the AGN jets. The minimum $t_{\rm cool}/t_{\rm ff}$ ratio is elevated. In cases where star formation cannot consume the cold gas quickly enough and a rotationally-supported disk forms (e.g., the second cycle), this ratio plateaus because the AGN self-regulates through the cold disk (see LB14a for details). As soon as the cold gas vanishes, which turns off AGN feedback, the ICM starts to cool. As one would expect, precipitation begins again after a period $\sim t_{\rm cool}$, marking the beginning of the next cycle.

The evolution of the star formation rate (SFR) is very similar to that of the average $\dot{M}_{\rm SMBH}$, while the history of the total amount of cold gas basically mirrors the SFR. 

Both $t_{\rm cool}$ and the minimum $t_{\rm cool}/t_{\rm ff}$ ratio indicate the general cooling properties of the ICM, but do not directly translate to the mass deposition rate -- i.e., the rate at which the ICM is cooling into the cold phase. Since an AMR code does not trace the history of fluid elements, we can only estimate this mass deposition rate $\rm \dot{M}_{\rm cooling}$. We can measure the changing rate of the total amount of cold gas $\rm \dot{M}_{\rm cold}$ which is related to $\rm \dot{M}_{\rm cooling}$:
\begin{equation}\label{eq:Mdot}
\dot{M}_{\rm cold} = \dot{M}_{\rm cooling} - \dot{M}_{\rm SMBH} -  SFR  - \dot{M}_{\rm heated}\; ,
\end{equation}
where the last term includes many physical processes that can reduce the amount of cold gas but are not strictly measurable in the simulations: shock heating, ram pressure stripping (of the cold clumps as they move through the hot ICM), Kelvin-Helmholtz instability (at the interface between cold clumps and the ICM) and SN heating. The first few processes are only significant when a large amount of cold clumps are present. Besides star formation, the depletion of the cold gas is dominated by SN heating.  

In an ideal situation where the thermal energy from SN feedback heats the cold gas exactly to the ICM temperature, we can write
\begin{equation}
\dot{E}_{\rm SN}=\epsilon_{\rm SNe}{\rm SFR}\times c^2=\frac{\dot{M}_{\rm SN heated}}{\mu m_p}\times \frac{3}{2} k_b T_{\rm ICM},
\end{equation}
where $k_b$ is the Boltzman's constant and $\epsilon_{\rm SNe} =10^{-5}$ (Section~\ref{sec:method_3}). Taking $T_{\rm ICM}=3$ keV and assuming a typical delay of 20 Myr between star formation and SN feedback, we have 
\begin{equation}
\dot{M}_{\rm SN heated}(t)=1.23\times {\rm SFR}(t-20 {\rm~Myr})\; .
\end{equation}

In reality, SN might overheat a cell to higher temperatures than the ICM temperature, resulting in an energy ``waste.''  More often, it under-heats the cell such that the temperature goes up to only $\sim 10^6 K$ or less where the cooling time is very short, and the cell quickly radiates the energy away and remains cold. Therefore, the calculation above (assuming optimum heating) likely overestimates the actual $\dot{M}_{\rm SN heated}$ in the simulation and should be taken as an upper limit.

\begin{figure*}[ht]
\begin{center}
\includegraphics[scale=.67,trim=0.9cm 0.8cm 0cm 0.2cm, clip=true]{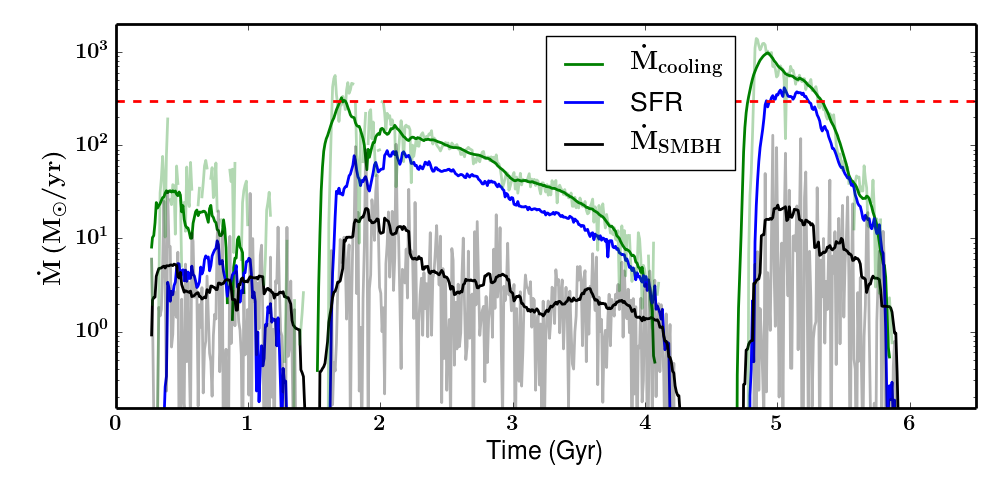}
\caption{The estimated mass deposition rate $\rm \dot{M}_{\rm cooling}$ (light green), the SFR (blue) and the SMBH accretion rate $\rm \dot{M}_{\rm SMBH}$ (grey) as a function of time. The solid green line and the black line show $\rm \dot{M}_{\rm cooling}$ and $\rm \dot{M}_{\rm SMBH}$ averaged over 200 Myr, respectively. The dashed red line ($300~\mathrm{M}_\odot\,\mathrm{yr}^{-1}$) shows the classical cooling flow prediction for $\rm \dot{M}_{\rm cooling}$ for the Perseus cluster.
\label{fig:Mdot_all}}
\end{center}
\end{figure*}

Approximating $\dot{M}_{\rm heated}$ as $\dot{M}_{\rm SN heated}$, we can rewrite Equation~\ref{eq:Mdot} as 
\begin{equation}
\dot{M}_{\rm cooling} \approx \dot{M}_{\rm cold} +  \dot{M}_{\rm SMBH} +  {\rm SFR} + \dot{M}_{\rm SN heated}\; .
\end{equation}
Figure~\ref{fig:Mdot_all} shows the estimated mass deposition rate $\dot{M}_{\rm cooling}$ as a function of time. Again, $\dot{M}_{\rm cooling}$ shows a similar trend to the SFR and $\dot{M}_{\rm SMBH}$, but peaks about 20 Myr earlier at the beginning of every cycle. As a reference, the red dashed line indicates the classical cooling rate of Perseus, which is recovered in our previous pure cooling flow simulation without AGN feedback \citep[][LB14a]{P1}. $\dot{M}_{\rm cooling}$ is significantly suppressed compared with the classical cooling flow prediction.

\begin{figure}
\begin{center}
\includegraphics[scale=.495,trim=1.1cm 0cm 0cm 0cm, clip=true]{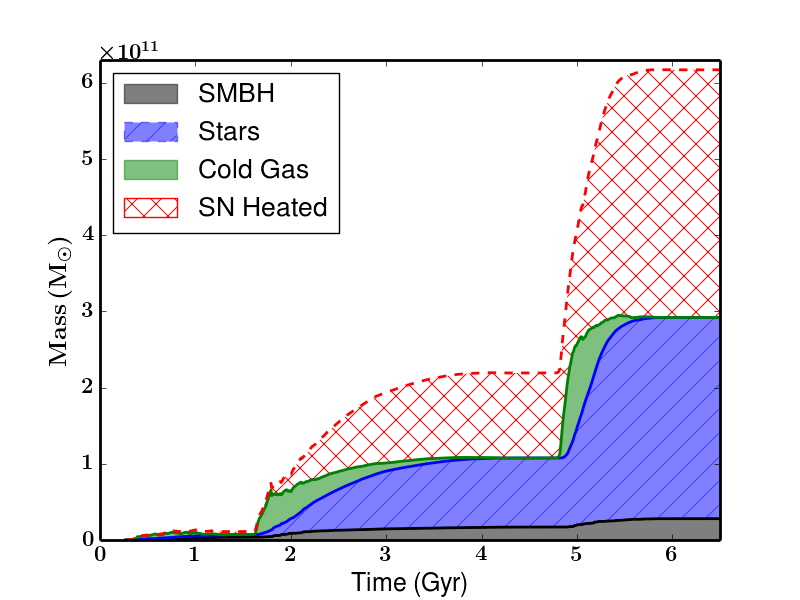}
\caption{The integrated amount of gas that is processed by the SMBH (grey), the gas that has formed stars (blue), the cold gas that exists in the system (green), and the estimated amount of cold gas that has been heated by SN feedback (red). 
\label{fig:TotalMass}}
\end{center}
\end{figure}

Over a period of 6.5 Gyr, about $6\times 10^{11}\ M_{\odot}$ of gas has cooled in the simulation, with roughly half ($\lesssim 3 \times 10^{11} M_{\odot}$) going into forming stars and the other half heated by supernovae (Figure ~\ref{fig:TotalMass}). The amount of gas that has been processed by the SMBH is $\rm 2.8\times 10^{10} M_{\odot}$.

\begin{figure*}
\begin{center}
\includegraphics[scale=.305,trim=0.4cm 0cm 0.6cm 0cm, clip=true]{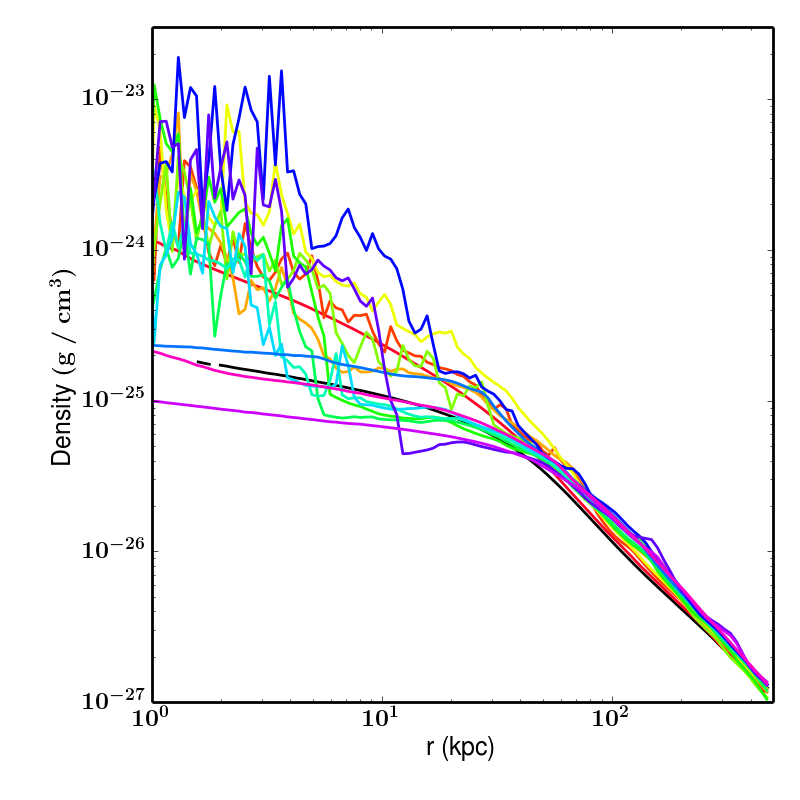}
\includegraphics[scale=.305,trim=0.4cm 0cm 0.6cm 0cm, clip=true]{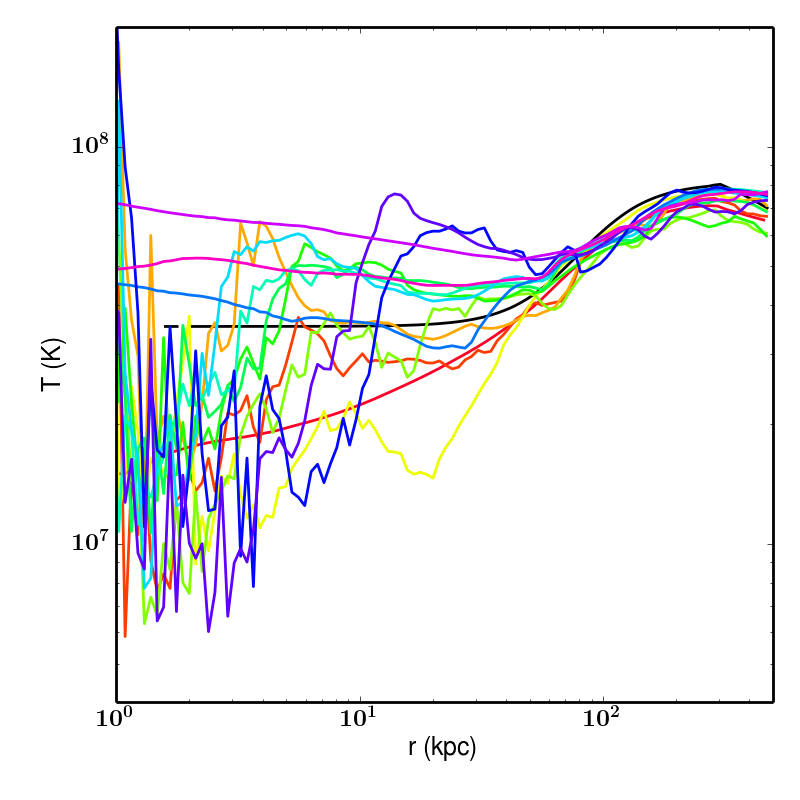}
\includegraphics[scale=.305,trim=0.4cm 0cm 0.6cm 0cm, clip=true]{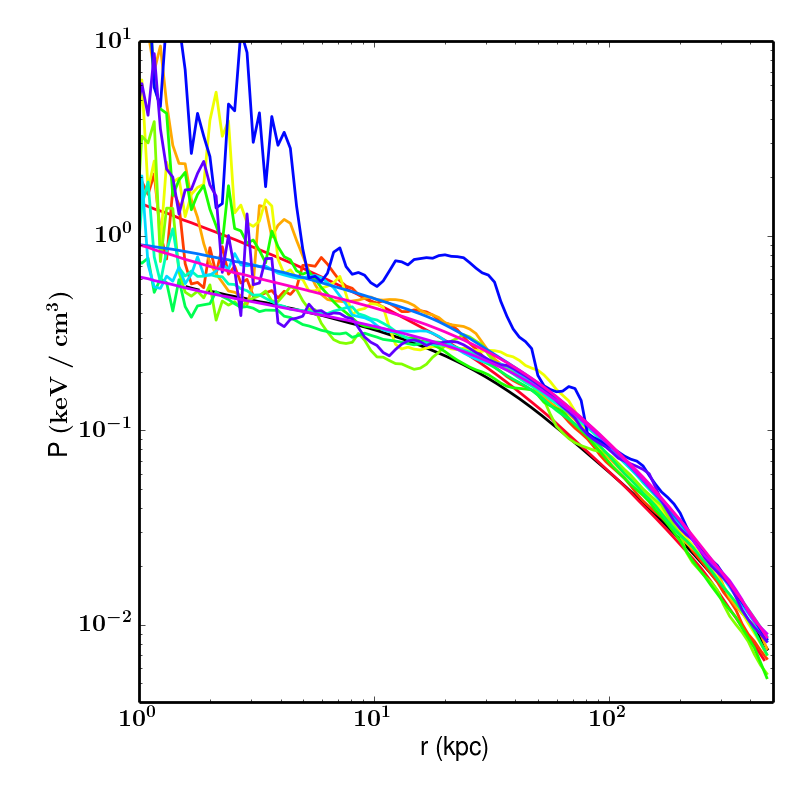}\\
\caption{Gas density, temperature and pressure profiles weighted by the X-ray emissivity. Profiles are plotted every 500 Myr since the AGN is first turned on, following the same color scheme as Figure ~\ref{fig:5in1}. The black line shows the initial condition at $t=0$. The profiles bounce around the initial condition and a cool-core appearance is maintained.
\label{fig:rainbow}}
\end{center}
\end{figure*}

\begin{figure}
\begin{center}
\includegraphics[scale=.5,trim=0.2cm 0.4cm 0cm 0cm, clip=true]{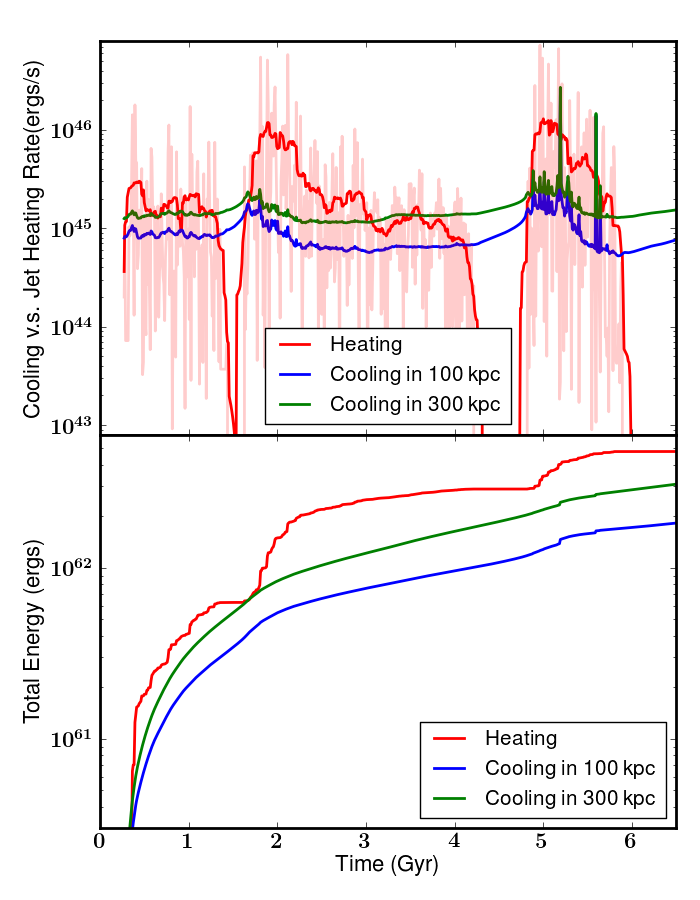}
\caption{Top: the AGN jet power ($\sim$the heating rate, shown in pink) and the total ICM cooling rate in the central $r<100$ kpc (blue) and $r<300$ kpc (green). The red line is the AGN heating rate averaged over 200 Myr. Bottom: The accumulated amount of heating from the AGN and the total cooling loss in the central $r<100$ kpc (blue) and $r<300$ kpc (green) of the cluster. 
\label{fig:Heating}}
\end{center}
\end{figure}

Figure ~\ref{fig:rainbow} shows that the radial profiles of the ICM density, temperature and pressure exhibit variations with time, but (outside of about 10 kpc) they never deviate much from the initial conditions and a cool-core appearance is preserved throughout the simulation (in particular there is a central core of gas which is cooler than the bulk of the cluster). The ICM cooling is balanced by AGN heating in a dynamical way: the jet power overwhelms cooling during the first half of every AGN cycle; it gradually declines until cooling start to dominate and eventually leads to precipitation which triggers the next cycle (top panel of Figure~\ref{fig:Heating}). At the end of the simulation, roughly 2/5 of the total jet energy is deposited within 100 kpc, and 3/5 within 300 kpc (bottom panel of Figure~\ref{fig:Heating}). These numbers are comparable to those found in the AGN-only simulations (LB14a).

We have neglected the heating from SNe in Figure~\ref{fig:Heating}. SN feedback heats the local cold gas effectively, but its impact on the hot ICM is minimal. Our SN feedback recipe corresponds to $10^{51}$ erg thermal energy output for every $180 M_{\odot}$ of stars formed, or 
\begin{equation}\label{eq:E_SN}
\dot{E}_{\rm SN}=1.76\times 10^{41} \big(\frac{\rm SFR}{1\ \rm M_{\odot}/yr}\big) {\rm ~erg}\,{\rm s}^{-1} \;.
\end{equation}
The average SFR in our simulation is $\rm 40 \, M _{\odot}$ yr$^{-1}$, so the average heating rate from SNe is $<\dot{E}_{\rm SN}>=7\times 10^{42} \,$ erg s$^{-1}$, three orders of magnitude lower than the AGN heating rate.

\subsection{Other Simulations}\label{sec:results_3}
In this section, we discuss the other simulations that we have performed in order to understand the role that individual physical processes play (Section~\ref{sec:othersims_2a}: AGN feedback; Section~\ref{sec:othersims_2b}: stellar feedback; Section~\ref{sec:othersims_2c}: self-gravity), and to study how the choices of simulation parameters affect the results.

\subsubsection{No AGN feedback}\label{sec:othersims_2a}
As a sanity check, we perform a simulation with star formation and stellar feedback but without AGN feedback. As expected, a classical cooling flow develops within a few hundred Myr and stars are forming at hundreds of $\rm M_{\odot}$ yr$^{-1}$. There is noticeable heating from SNe in the very center of the cluster as compared to a run without star formation or AGN feedback. However, the heating rate (estimated using equation \ref{eq:E_SN}) is only a few times $10^{43}\,$ \rm erg s$^{-1}$, far below the ICM cooling rate. 

This confirms that although star formation and stellar feedback have a significant impact on the cooling-AGN feedback cycle, AGN feedback still plays the major role in heating the ICM.

\subsubsection{No Stellar Feedback}\label{sec:othersims_2b} 
Prior to our main simulation, we also performed runs that include star formation but no stellar feedback. The results are very similar to our standard run. This is simply because stellar feedback is only roughly as effective as star formation itself in reducing the amount of cold gas in the system (Figure~\ref{fig:TotalMass}). The primary difference we observe is that without stellar feedback, the star formation rate is higher than in our standard run, as one would expect \citep{Tasker2006}.

\subsubsection{The Effect of Self-gravity}\label{sec:othersims_2c}
We have also performed simulations to test the importance of including the self-gravity of the gas. The gravitational potential is dominated by the dark matter on large scales and by the stellar component of the BCG in the innermost $\sim 10$ kpc, while the contribution from the ICM is negligible \citep{P1}. This is the main reason why previous simulations focusing on the AGN feedback usually ignore the self-gravity of the gas \citep[e.g.,][LB14a \& b]{Gaspari2011}. However, self-gravity may assist the development of gravitational instabilities of the cold disk, which can help redistribute angular momentum and increase the accretion rate onto the SMBH \citep{Lodato2007}. 

Our first test run includes AGN feedback and self-gravity but no star formation. The cluster evolution within the first 1.5 Gyr is very similar to the standard run discussed in LB14a without the self-gravity of the gas. The cold gas settles into a rotating disk by $t=2$ Gyr. Self-gravity triggers gravitational instabilities, and the disk shows structures that resemble the spiral arms of disk galaxies. The accretion rate is higher compared with LB14a because the spiral-mode instabilities transport angular momentum, and enhances accretion. Thus the core temperature is also more elevated. The amount of cold gas is reduced, but the ICM is still cooling. The SMBH accretion rate is enhanced due to self-gravity, but it is still not enough to consume all the cold gas in the cluster center by the end of the simulation (within a few Gyr). Therefore, the problem of a long-lived massive cold disk persists, and no clump formation is seen after $t=2$ Gyr in the over-heated cluster core. This means that self-gravity may alleviate the problem but is not the key to solving it. 

The other test we have performed is a simulation with all the ingredients included in our standard run except self-gravity. We find that the absence of self-gravity does not significantly alter the results. The cluster still experiences AGN outburst cycles. The main difference is that without self-gravity, the SFR is less bursty than in our standard run, peaking only at $\rm \sim 150\, $M$_{\odot}$\ yr$^{-1}$.  This is likely because self-gravity can affect the density distribution of the star-forming clouds and speeds up the initial collapse \citep{Kritsuk2011, Lee2014}. 
Another difference is that without the assistance of self-gravity, the accretion of the cold disk onto the SMBH is less efficient. Thus, within 6.5 Gyr the cluster only experiences two major cycles rather than the three seen in our standard run. The cold disk also does not completely vanish (but is significantly reduced in mass) at the end the first cycle.

\subsubsection{SMBH feedback efficiency}\label{sec:othersims_2d}
Previous simulations have found that due to self-regulation, AGN feedback with a wide range of feedback efficiencies ($\epsilon=0.1\%-1\%$) produces similar results \citep[][LB14a]{Gaspari2012}. This changes when star formation is included.

Although star formation and SMBH feedback evolve hand-in-hand (Figures~\ref{fig:5in1} and~\ref{fig:Mdot_all}) as the ICM goes through cycles of cooling and heating, they also compete with each other at any given time as they share the same fuel -- cold dense gas. In fact, as discussed earlier, the amount of cold gas that feeds star formation is more than an order of magnitude higher than that accreted onto the SMBH (see Section~\ref{sec:results_2} or Figure~\ref{fig:SFR_SMBH}). 

In LB14a, we chose $\epsilon=0.1\%$ over $\epsilon=1\%$ as our preferred value (and thus used this value in the standard run) mainly because a feedback efficiency of $\epsilon=1\%$ slightly overheats the core at late times. In the standard simulation discussed here, we use $\epsilon=1\%$ because with most of the cold gas going into forming stars, a low AGN feedback efficiency is no longer sufficient. 

\begin{figure}
\begin{center}
\includegraphics[scale=.53]{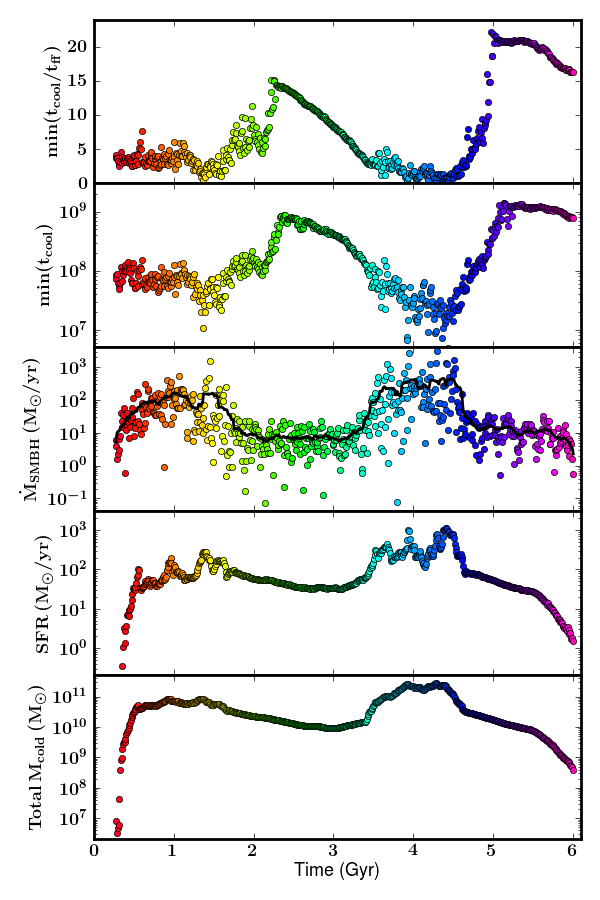}
\caption{The time evolution of the minimum $t_{\rm cool}/t_{\rm ff}$ ratio, the minimum cooling time $t_{\rm cool}$, the SMBH accretion rate $\rm \dot{M}_{\rm SMBH}$, the SFR, and the total amount of cold gas in the simulation with a lower feedback efficiency $\epsilon=0.1\%$. The data is sampled every 10 Myr and the color scheme is the same as Figure ~\ref{fig:5in1}. The low efficiency results in too much star formation and too much cold gas.
\label{fig:5in1_low}}
\end{center}
\end{figure}

Figure~\ref{fig:5in1_low} shows the cluster evolution for the simulation with $\epsilon=0.1\%$ and everything else kept the same as the standard run. AGN feedback still manages to prevent the development of a classical cooling flow, and the cluster still shows somewhat cyclical behaviors similar to the standard run. This is not surprising since the average $\dot{M}_{\rm SMBH}$ here is roughly 10 times that in the standard run, making up for the lower efficiency, manifesting again the self-regulating nature of AGN feedback.  However, the total mass of the cold gas in the system is almost always above $10^{10} M_{\odot}$ and the SFR is also too high, with an average of 120 $\rm M_{\odot}$\ yr$^{-1}$. Thus a low feedback efficiency of $\epsilon=0.1\%$ can be ruled out.

\subsubsection{Star Formation Prescription}\label{sec:othersims_2e}
There are many parameters related to the star formation routine described in Section~\ref{sec:method_3}. The values we have chosen for our simulations are typically used in simulations of star forming galaxies that produce results roughly consistent with the observations \citep[e.g.,][]{Tasker2006, Hummels}. A systematic study of varying the values for all those parameters is beyond the scope of this work. Here we briefly discuss the results from a test run with a lower minimum mass for star particles ($m_{*,\rm min}=5\times 10^4 \ \rm M_{\odot}$, 1/20 of the value used in our standard run). 

Because star particles are easier to form at the beginning of the cooling cycle, more cold gas goes into forming stars and less gets accreted onto the SMBH, which leads to less energetic AGN outbursts. As a result, less condensation is triggered by the jets, and the first cycle is shorter (lasting for less than 1 Gyr) and gentler than that in the standard run. After global turbulence is built up by $t=1$~Gyr, the cluster behaves more similarly to our standard run. The range of SFR and the minimum $t_{\rm cool}/t_{\rm ff}$ ratio stay roughly the same, except that the tail of the SFR distribution extends a bit further, as one might expect. As discussed in Section~\ref{sec:results_2}, the direct effect of star formation and stellar feedback on the hot ICM is negligible. We do not expect the results to change qualitatively when varying star formation parameters because the battle is ultimately between ICM cooling and AGN feedback. However, due to the complex relations between cooling, star formation and AGN feedback, the exact choice of those parameters could have an effect on the strength and duration of the long term ($\gtrsim$ Gyr) AGN cycles.

\section{Discussion}\label{sec:discussion}
In this section, we compare our results with observations and other simulations done previously (Section~\ref{sec:discussion_1}), and we discuss the possible impact of other physical processes on the simulation results (in Section~\ref{sec:discussion_2}).

\subsection{Comparison with Observations}\label{sec:discussion_1}

A number of recent simulations show that gas condensation occurs when the minimum $t_{\rm cool}/t_{\rm ff}$ ratio of the system drops below a certain threshold. This threshold is found to be $\sim 1$ in simulations with idealized heating in \citet{McCourt12}, and $\sim 10$ in a spherical geometry \citep{Sharma2012, Sharma2015}, which is confirmed in \citet{Gaspari2012} with more realistic AGN feedback. LB14b finds that the value of this critical ratio can vary between $\sim 3-10$ depending on the strength of the perturbation and the way AGN jets interact with the ICM (dredging up gas with shorter $t_{\rm cool}$ to larger radii where $t_{\rm ff}$ is longer).

\begin{figure}
\begin{center}
\includegraphics[scale=.485,trim=0.8cm 0.1cm 0cm 0cm, clip=true]{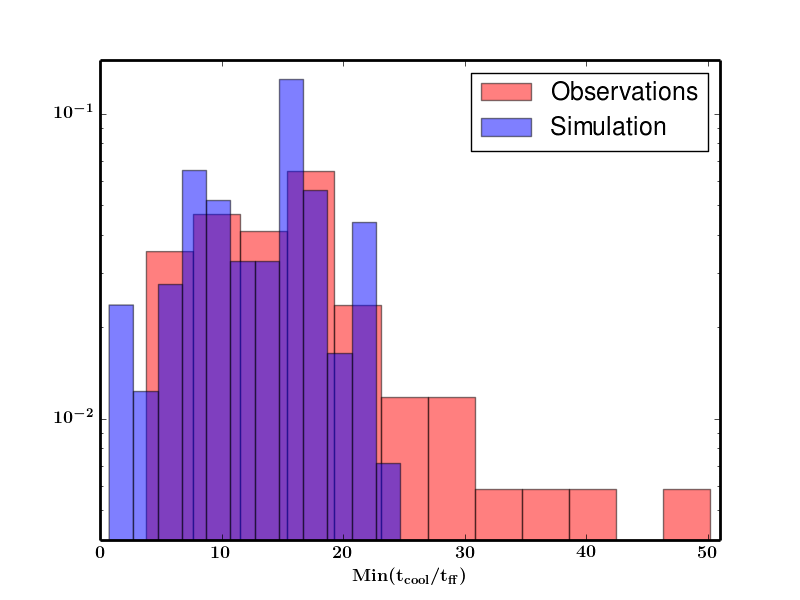}
\caption{The normalized distribution of the minimum $t_{\rm cool}/t_{\rm ff}$ ratio at times when cold gas exists in our standard run (blue) compared with the observations (red) for precipitating systems (data from \citet{MarkMegan2014}).
\label{fig:ratio_hist}}
\end{center}
\end{figure}

Figure~\ref{fig:ratio_hist} shows the distribution of the minimum  $t_{\rm cool}/t_{\rm ff}$ ratio for the times when cold gas exists in the system. Precipitation can happen when the minimum $t_{\rm cool}/t_{\rm ff}$ ratio is as low as $\sim 1$ (only very briefly) when the AGN is off, and as high as $\sim 20$ at the peak of the AGN outburst when the ICM is most perturbed. Over-plotted in Figure~\ref{fig:ratio_hist} are the observations of cool-core clusters with $\rm H\alpha$ detection. The bulk of the two distributions are consistent with each other. The observed distribution shows a longer tail at the high end and a deficit at the very low end where the minimum $t_{\rm cool}/t_{\rm ff}$ ratio is $\lesssim$ 3. This discrepancy may be due to mergers and/or cosmological infall, effects that are not included in our isolated simulations.

\begin{figure}
\begin{center}
\includegraphics[scale=.465,trim=0.7cm 0.5cm 0cm 0cm, clip=true]{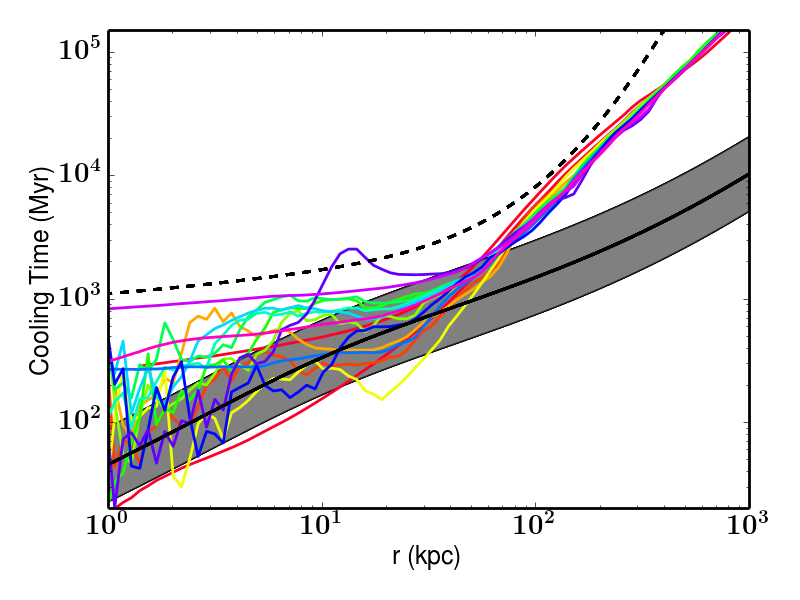}
\caption{The cooling time $t_{\rm cool}$ profile sampled every 500 Myr since the AGN was first turned on. The color scheme is the same as previous figures (from red-early to purple-late). The solid black line is $10t_{\rm ff}$ and the grey shaded region is $5t_{\rm ff}$ to $20t_{\rm ff}$. The dashed black line is the conductively balanced solution and is discussed in Section~\ref{sec:discussion_2}. \label{fig:Mark_time}}
\end{center}
\end{figure}

Observationally, in precipitating systems the radial profile of $t_{\rm cool}$ is usually above $\sim 10~t_{\rm ff}$ and never below $5~t_{\rm ff}$ \citep[Figure 1 of][]{Mark2014Nature1}. Our simulation generally agrees with this finding (Figure~\ref{fig:Mark_time}). Note that we plot only the initial $t_{\rm ff}$ for clarity, but at later times the stars formed in the simulation also add mass and thus slightly steepen $t_{\rm ff}$. Despite this effect, our system is still found to have relatively low $t_{\rm cool}$ (between 5 and 10 $t_{\rm ff}$) more often than the observations seem to indicate. This may be partially due to the difference in the way $t_{\rm cool}$ is measured in simulations and in observations. For example, the yellow line that shows a dip at $r\sim 20$ kpc corresponds to the off-centered cooling at the beginning of the second cycle. Observationally, this off-centered low entropy blob could be identified as the X-ray center and thus change the spherically-averaged $t_{\rm cool}$.

\begin{figure*}[ht]
\begin{center}
\includegraphics[scale=.7,trim=0cm 0.3cm 0cm 0cm, clip=true]{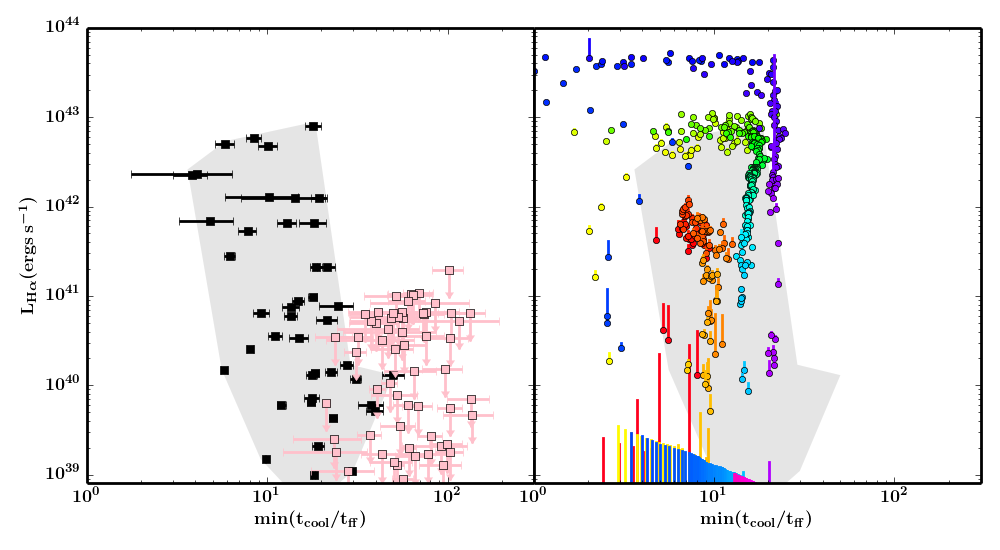}
\caption{Left: the observed $\rm H\alpha$ luminosity $L_{\rm H\alpha}$ vs. the minimum $t_{\rm cool}/t_{\rm ff}$ ratio in \citet{MarkMegan2014}. The black squares are the systems with $\rm H\alpha$ detection, and the pink squares are the ones with only upper limits. Right: the estimated $L_{\rm H\alpha}$ vs. the minimum $t_{\rm cool}/t_{\rm ff}$ ratio in the standard run. The data is sampled every 10 Myr and the color scheme is the same as Figure ~\ref{fig:5in1} (from red-early to purple-late). The circles correspond to the $L_{\rm H\alpha}$ associated with star forming regions and the vertical bars indicate the estimated maximum contribution from collisional ionization. The shaded grey region in both panels denotes the range that the observed precipitating systems cover. 
\label{fig:Halpha}}
\end{center}
\end{figure*}

As is shown in Section~\ref{sec:results_2}, star formation is closely linked to ICM cooling, which can be characterized with the cooling time, $t_{\rm cool}$. Observations confirm that they are indeed correlated: $\rm H\alpha$ luminosity $L_{\rm H\alpha}$ tends to be higher in systems with lower central entropy or smaller minimum $t_{\rm cool}/t_{\rm ff}$ ratio; it is often undetected if the ICM is hotter than some threshold (e.g. $\rm K_0 \gtrsim 30 \, keV \, cm^2 $ or $t_{\rm cool}/t_{\rm ff} \gtrsim 20$) \citep{Cavagnolo2008, McDonald10, Rawle2012, MarkMegan2014}. We show in Figure~\ref{fig:Halpha} the relation between the minimum $t_{\rm cool}/t_{\rm ff}$ ratio and the estimated $L_{\rm H\alpha}$ in our simulation. To obtain $L_{\rm H\alpha}$ from star forming regions, we adopt the commonly used linear relationship: $L_{\rm H\alpha}$(ergs s$^{-1}$) = $1.26\times 10^{41}$  SFR\ (M$_{\odot}$\ yr$^{-1})$ \citep{Kennicutt1998}. Besides photoionization, some $\rm H\alpha$ emission may also result from collisional ionization and the central AGN. For simplicity, we only calculate the emission from collisional ionization through interpolation of a table generated with Cloudy 13.00 \citep{Cloudy} assuming coronal equilibrium. This gives an upper limit because we are double-counting the photoionized regions. The contribution from collisional ionization is shown as the vertical bars and is much lower than photoionization most of the time (smaller than the size of the circles).

Our system generally occupies the same area on the $L_{\rm H\alpha}$-min$(t_{\rm cool}/t_{\rm ff})$ plot as the observed cool-core clusters with detected $\rm H\alpha$ emission \citep[data from][]{MarkMegan2014}. There appear to be more observed systems with relatively low $L_{\rm H\alpha}$ ($\sim 10^{40}$ erg s$^{-1}$) and high min$(t_{\rm cool}/t_{\rm ff})$ ($\sim 10-30$) than our simulation for a number of reasons. First, min$(t_{\rm cool}/t_{\rm ff})$ is measured with narrower radius bins in our simulations than most observations, which tends to give a smaller value. Second, we do not include mergers in our simulations, which could trigger precipitation when min$(t_{\rm cool}/t_{\rm ff})$ is higher ($\sim 20-30$). Third, we only simulate one system based on Perseus, a relatively massive cluster. It is possible that the area with low $L_{\rm H\alpha}$ ($\sim 10^{40}$  erg s$^{-1}$) will be more populated if we add more simulation data with systems of various different sizes. Lastly, the star formation rate is likely too high in our simulations, which is a common issue in numerical simulations that only include thermal feedback from SNe \citep[e.g.][]{Katz1992}. A more realistic SNe feedback model with kinetic feedback \citep[e.g.,][]{Dalla2008, Christine} or including other physical processes such as cosmic rays \citep{Munier} and magnetic fields \citep{Loo} would likely suppress star formation and lower the SFR. 

\begin{figure}
\begin{center}
\includegraphics[scale=.48,trim=0.5cm 0.1cm 0cm 0cm, clip=true]{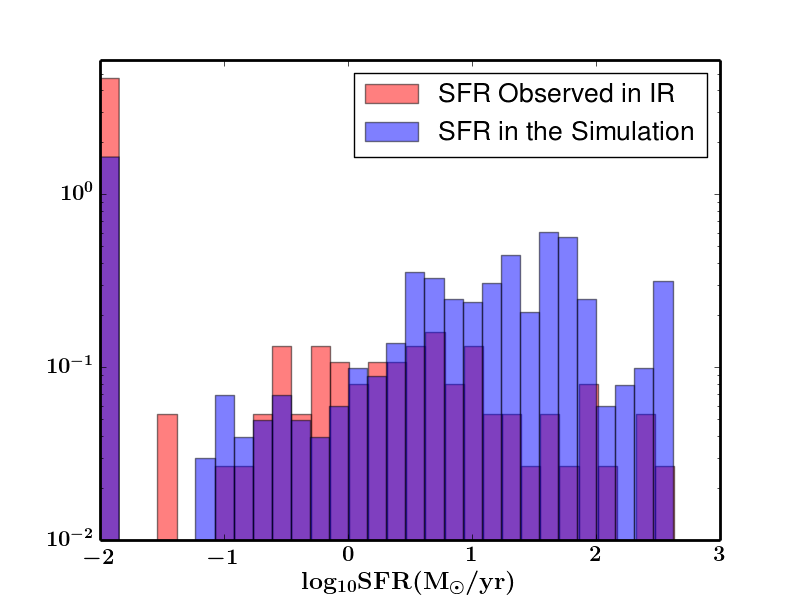}
\caption{The normalized distribution of the SFR in our standard run (blue) compared with the IR observations (red) of the BCGs in the ACCEPT sample (\citet{Hoffer2012}). For clarity, we assign a single value of $10^{-2}\rm M_{\odot}/yr$ to all the upper limits in the observations and the times with no SFR in our simulations (the leftmost bar in the distribution).
\label{fig:SFR_hist}}
\end{center}
\end{figure}

The peak of the $\rm H\alpha$ luminosity ($L_{\rm H\alpha} > 10^{43}$ erg s$^{-1}$) in our simulation appears to exceed all the observed data in Figure~\ref{fig:Halpha}, partly due to the reasons discussed above, and partly due to the incompleteness of the $\rm H\alpha$ sample. If we compare the SFR in the simulation with the SFR observed in the infrared (IR), we find that the two cover roughly the same range (Figure~\ref{fig:SFR_hist}). 
The range of the SFR in our idealized simulation is also comparable to that found at late times in a cosmological simulation in \citet{Dubois2010}.

Figure~\ref{fig:Halpha} also shows that there is no linear anti-correlation between $L_{\rm H\alpha}$ and the minimum ratio, as one might naively expect. This is because star formation rate is correlated with the amount of cold gas, but not the rate at which gas is cooling (which is correlated with $t_{\rm cool}/t_{\rm ff}$ ratio). Within each cycle, there is a significant delay up to 1-2 Gyr between when the ICM cooling rate is suppressed (characterized as the increase of the $t_{\rm cool}/t_{\rm ff}$ ratio) due to AGN heating and when SFR starts to decline at the end of the cycle due to the consumption of the cold gas. As a result of this hysteresis, there is no linear relation between SFR and the cooling rate (or the $t_{\rm cool}/t_{\rm ff}$ ratio), but they are related in that star formation only occurs when the $t_{\rm cool}/t_{\rm ff}$ ratio is below some critical value.

\begin{figure}
\begin{center}
\includegraphics[scale=.55,trim=0.2cm 0.4cm 0cm 0cm, clip=true]{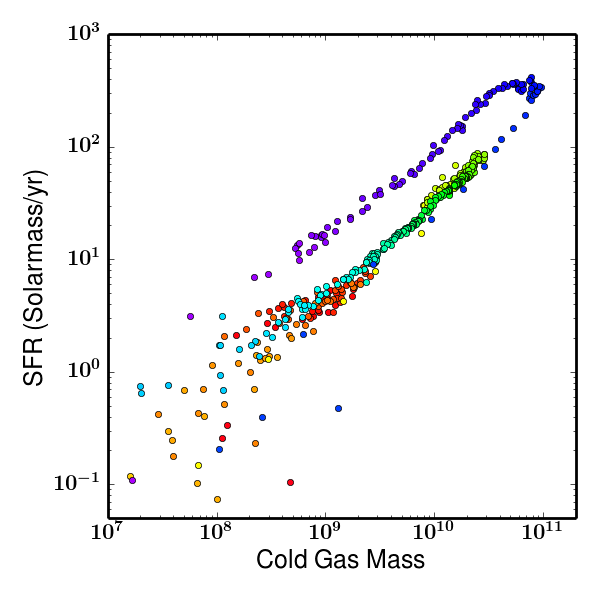}
\caption{The correlation between SFR and the total amount of cold dense gas ($T<10^5 K$ and number density above 10 cm$^{-3}$) in the system. The data is sampled every 10 Myr and the color scheme is the same as Figure ~\ref{fig:5in1}. 
\label{fig:SFR_coldgas}}
\end{center}
\end{figure}

Figure ~\ref{fig:SFR_coldgas} shows the relationship between SFR and the total amount of cold dense gas (with $T<10^5$ K and number density above 10 cm$^{-3}$). A tight correlation spanning a few orders of magnitude is maintained over the course of the simulation. The slope is generally consistent with what has been found between the SFR and the inferred $H_2$ mass for BCGs \citep{ODea2008}, and for star forming galaxies when comparing SFR and the amount of dense molecular gas \citep{GaoYu2004}. A more detailed comparison with the observations can further constrain the star formation efficiency $\epsilon_{\rm SF}$ used in our model. We leave this to future work.

\begin{figure*}
\begin{center}
\includegraphics[scale=.69,trim=0.2cm 0.4cm 0cm 0cm, clip=true]{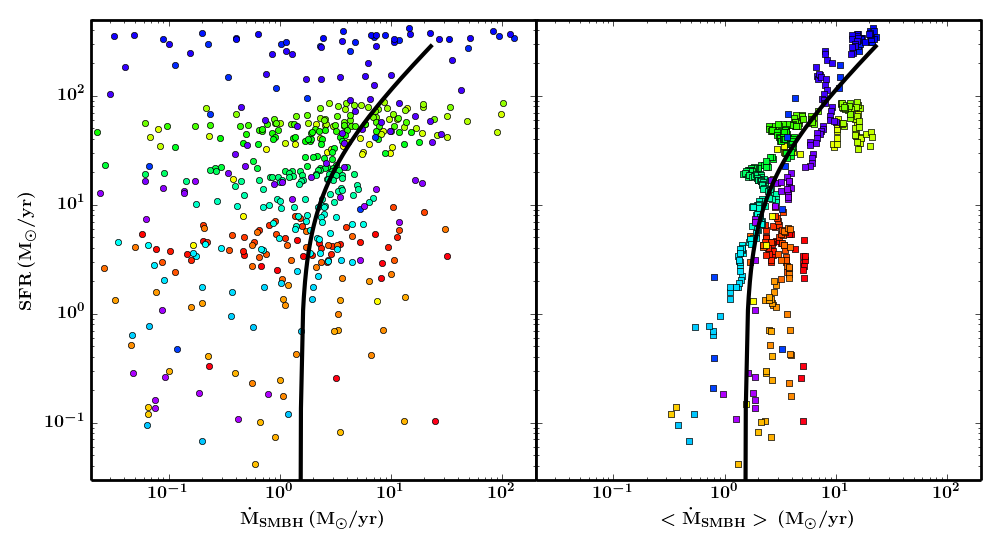}
\caption{Left: SFR vs. instantaneous SMBH accretion rate $\rm \dot{M}_{\rm SMBH}$ (dots). No clear correlation is seen due to the large scatter. Right: SFR vs. $\rm <\dot{M}_{\rm SMBH}>$, the SMBH accretion rate averaged with a moving window of 200 Myr (squares). The color scheme is the same as previous figures (from red-early to purple-late). In both panels the solid black line shows the best linear fit to the data on the right: $SFR=13.4\times <\dot{M}_{\rm SMBH}>-20.6$.
\label{fig:SFR_SMBH}}
\end{center}
\end{figure*}

Unlike SFR, observations have failed to find a clear correlation between the radio power of the AGN and the cooling properties of the ICM \citep{Mittal2009} or the amount of cold gas in the system \citep{Salome2003}. It is possible that the radio power does not correlate with the actual mechanical power of the AGN, but even if it does \citep{Cavagnolo2010}, our simulation shows that a clear correlation would be surprising simply due to the large fluctuations in the instantaneous AGN power: as Figure~\ref{fig:5in1} and \ref{fig:Mdot_all} show, the history of the SFR is rather smooth within each cycle, but the AGN power varies on very short timescales ($\leq$ 10 Myr, the sampling time interval), because the accretion region is very small and the flow of the cold gas is not smooth. The left panel of Figure~\ref{fig:SFR_SMBH} shows the SFR plotted against the instantaneous $\dot{M}_{\rm SMBH}$, and no correlation can be observed. However, the average AGN power does correlate with ICM cooling and star formation as discussed in Section~\ref{sec:results_2}. The right panel of Figure~\ref{fig:SFR_SMBH} shows the relation between SFR and the average $\dot{M}_{\rm SMBH}$ (calculated by smoothing $\dot{M}_{\rm SMBH}$ with a 200 Myr moving window, corresponding to the black line in the third panel of Figure~\ref{fig:5in1}). The solid black line is the best linear fit:

\begin{equation}
\rm SFR=13.4 \big<\dot{M}_{\rm SMBH}\big>-20.6\;, 
\end{equation}
with both SFR and $\big<\rm \dot{M}_{\rm SMBH}\big>$ measured in $\rm M_{\odot}/yr$. The root mean square (RMS) error is 55. When averaged with a shorter moving window of 50 Myr, the RMS increases to 70.

We also over-plot the best-fit line on the left panel for guidance, where the data shows large scatter and appears to be distributed widely, except for the lower right corner of the panel. This suggests that although $\dot{M}_{\rm SMBH}$ varies a lot with time, objects with very high $\dot{M}_{\rm SMBH}$ and very low SFR should be very rare.

Significant AGN variability on short timescales has been used to explain the lack of a strong correlation between AGN activity and star formation in galaxies, as one would expect from the $M-\sigma$ relation \citep[e.g.,][]{Hickox2014, Thacker2014}. Our simulation confirms that even though the star formation rate does not appear to correlate with the instantaneous $\dot{M}_{\rm SMBH}$, a strong relationship does exist between SFR and the average $\dot{M}_{\rm SMBH}$. If we assume that the SMBH growth rate is $\sim 1\%$ of $\dot{M}_{\rm SMBH}$, then the ratio of the mass added to the SMBH to that added to stars is approximately $10^{-3}$, consistent with the observed $M-\sigma$ relation \citep{Msigma}.

\subsection{Thermal Conduction and Other Physics}\label{sec:discussion_2}
One important piece of physics not included in our simulations is thermal conduction. Conduction can be a very efficient heating mechanism \citep{Rosner, Narayan2001}, but it is generally found that conduction fails to stem cooling in the innermost $\sim$ 20 kpc \citep{Ruszkowski2002, Voigt, Smith2013}, or it may be sufficient to balance cooling in some systems but not others with denser cores \citep[e.g.][]{Narayan2003, Parrish2009, RO10, Ruszkowski2011}. We plan to explore the effect of thermal conduction numerically in the context of AGN feedback in future work. Here we only qualitatively study its importance and try to address the question: does AGN feedback ever boost the core entropy to such a level that conduction can take over and turn a cool-core cluster into a non cool-core cluster \citep{Guo2009}?

Assuming the ICM is in hydrostatic equilibrium with a given gravitational potential, one can derive a family of quasi-steady solutions where radiative cooling is perfectly balanced by thermal conduction at every radius \citep{Guo2008, Voit2011}. If a cluster has an entropy profile above this solution, conduction will make the core isothermal; if it is below the solution, conduction will fail to balance cooling and a cooling flow would develop in the absence of other heating sources \citep{Bertschinger1986}. The dashed line in Figure~\ref{fig:Mark_time} shows a conductively balanced solution for a 30\% Spitzer conductivity \citep{Spitzer} with the inner boundary conditions:
\begin{equation}
n_e(0)=0.12 \ {\rm cm}^{-3}\;;
T(0)=T(1)= 4 \rm keV\;. 
\end{equation}

Our cluster spends most of the time below the conductively balanced solution. We caution that this solution does depend on the choice of the boundary conditions, and varying ($n_e(0)$, $T(0)$) within a reasonable range can shift the solution up and down by $50\%$. Still, given how briefly the cluster profile approaches or exceeds the conductively balanced line, it is unlikely that thermal conduction would have enough time to transform the cluster core. Therefore, our simulation suggests that even with the help of conduction, AGN feedback seems unable to boost a core-core cluster to a non cool-core cluster.

However, thermal conduction may reduce the amount of cold gas, and thus lower the star formation rate. On the other hand, the transition layer surrounding cold clumps may become wider due to conduction, resulting in an increase in the $\rm H\alpha$ luminosity. In addition, thermal conduction may help distribute the heat from the AGN jets more evenly, and it is almost certainly important in heating the outer parts of the cluster core ($\sim30-300$ kpc). Without conduction, by the end of our simulation, there is noticeable cooling around $100-300$ kpc (Figure~\ref{fig:rainbow}): the temperature has decreased and the density has increased. This is also shown in Figure~\ref{fig:Heating}: the separation between the energy loss rate due to cooling within 100 kpc and 300 kpc increases with time. Even though cooling is roughly balanced by AGN heating globally, without conduction the inner core gets slightly overheated (the average cooling rate within 100 kpc declines) while the outer core is still under heated (cooling within 300 kpc grows). This may also be why each burst of precipitation/AGN feedback is stronger than the previous one. 

We have already discussed the possible impact of mergers on the distribution of minimum $t_{\rm cool}/t_{\rm ff}$ ratio in Section~\ref{sec:discussion_1}. Merger-induced dynamical heating may have similar effects as conduction in heating the outer part of the cluster core around $100-300$ kpc, and thus reduced the burden on the AGN.

Another piece of physics missing in our simulations is magnetic fields. Even though with star formation we no longer see a massive cold disk that lasts indefinitely, our cluster still harbors a (still fairly massive) disk roughly half of the time in the simulation. This is higher than what observations seem to suggest \citep{McDonald10}. Magnetic fields may support the cold filaments against gravity \citep{Fabian2008} and delay the formation of a disk, and thus reduce the time the cluster spends in the disk phase. 

Magnetic fields and cosmic rays can also provide pressure support, particularly in the condensed gas. Figures~\ref{fig:5in1} and~\ref{fig:Mdot_all} show that gas condensation at the beginning of each cycle always happens in a catastrophic way where the amount of cold gas surges up drastically\footnote{Note that this could also be in part due to numerical overcooling at the edges of cold clouds \citep{Brighenti2015}}. Since the growth of individual cold clumps is mainly driven by the pressure difference (the cold clumps are typically under-pressured), including magnetic fields and cosmic rays in the simulation may slow down the growth and result in less bursty cooling and AGN feedback. In addition, both magnetic fields and cosmic rays could suppress star formation \citep{Munier, Loo}. We again leave this to future work.

\section{Conclusions}\label{sec:conclusion}

We have performed three-dimensional AMR simulations of an idealized, isolated cool-core galaxy cluster based on Perseus, focusing on the interplay between ICM cooling, AGN feedback and star formation over 6.5 Gyr. The momentum-driven AGN feedback is modeled with a pair of jets precessing at a small angle. The jets are powered by the accretion of cold gas in the close vicinity of the SMBH in the center of the cluster potential. We also include radiative cooling, self-gravity of the gas, star formation and stellar feedback in the form of thermal energy. We perform parallel simulations to tease apart the roles individual physical processes play, and compare our simulations with the observations. The key results of this work are summarized below.

\begin{enumerate}
\item The radiative cooling of the ICM, AGN feedback and star formation are tightly linked in cool-core clusters. Momentum-driven AGN jets trigger the low entropy gas to condense out of the ICM. Some of the condensed gas is accreted onto the SMBH to power the AGN jets, but much of it forms stars. While AGN heating elevates the entropy of the ICM in the cluster center, lengthening the cooling time, star formation and stellar feedback gradually consume and erode the cold gas (typically over 1-2 Gyr). When the cold gas vanishes, AGN feedback briefly turns off and allows the ICM to cool again, until gas with a short cooling time starts to precipitate, which triggers the next AGN outburst. Even though AGN feedback is the primary heat source for the ICM, star formation and stellar feedback are more effective in depleting the cold gas, thus regulating the long-term AGN cycles.

\item The range of star formation rate and the minimum $t_{\rm cool}/t_{\rm ff}$ ratio in the simulation generally agree remarkably well with the observations.  Our cluster does occasionally show a higher star formation rate and a lower $t_{\rm cool}/t_{\rm ff}$ ratio than observations indicate -- this slight discrepancy could be related to many factors: mergers and cosmological infall, more effective stellar feedback, as well as the other physical processes not included in our simulations, such as conduction and magnetic fields.

\item The instantaneous SMBH accretion rate in the simulations shows large variations on short timescales ($\leq$ 10 Myr). This explains why a clear correlation is not observed between the AGN radio power and star formation rate or the cooling properties of the ICM. Nevertheless, SMBH accretion and star formation are still related to each other, because both are directly linked to cooling. The star formation rate is tightly correlated with the SMBH accretion rate when the accretion rate is averaged over $\sim$ 200 Myr.

\item Star formation is proportional to the total amount of cold gas, not the rate at which the ICM is cooling. The ICM responds to AGN heating very quickly, but the decline of the star formation rate occurs over a much longer period of time. As a result, even though star formation only happens when the minimum $t_{\rm cool}/t_{\rm ff}$ ratio of the ICM is lower than some threshold ($\sim 20$ in our simulations), there is no linear correlation between the two. This is consistent with the observations of precipitating cool-core clusters. 

\item The main results of this paper (the cyclical behavior of the cluster, the distribution of the minimum $t_{\rm cool}/t_{\rm ff}$ ratio, the general range of star formation rates, etc) are fairly robust when we exclude stellar feedback, or the self-gravity of the gas, or if star formation parameters are changed. Though these changes do have an effect on the strength and duration of individual cycles, the ratio between SFR and the SMBH accretion rate, and the exact distribution of SFR. 

\item Due to its self-regulating nature, SMBH feedback can balance cooling with a wide range of feedback efficiencies, from 0.1\% to 1\% in numerical simulations without star formation. When star formation is included, however, even though a thermal balance is still achieved with a low feedback efficiency of 0.1\%, the star formation rate is unrealistically high and the system has too much cold gas for too long. Therefore, in the absence of other physics, a low SMBH feedback efficiency of 0.1\% can be ruled out.

\item AGN feedback balances ICM cooling in a dynamical way: during major AGN outbursts at the beginning of every cycle, there is more heating than cooling; as the cooling rate decreases and star formation consumes the fuel, heating yields to cooling until the beginning of the next cycle. Heating and cooling are balanced in an average sense, and a cool-core appearance is maintained despite the fluctuation. Our analysis shows that even with the help of thermal conduction, it is unlikely that AGN outbursts can make a cool-core isothermal. However, conduction may heat the outer parts of the cool-core at radii of roughly 100 kpc, where AGN heating becomes less effective. 
\end{enumerate}

Our numerical model including both momentum-driven AGN feedback and star formation not only solves the ``cooling flow problem'' but also reproduces the behavior of cool-core clusters that are largely consistent with observations. We do not claim, however, that we have found the best parameter set. Future work with a more systematic parameter study and a more extensive comparison with the observations can put better constraints on our model. A more sophisticated setup that includes mergers, conduction, and magnetic fields may also bring the model even closer to reality.

\acknowledgments

We acknowledge financial support from NSF grants AST-0908390, AST-1008134, AST-1210890, AST-1008454, NASA grants NNX12AH41G, NNX12AC98G, and ATP12-ATP12-0017, Hubble Theory Grant HST-AR-13261.01-A,  as well as computational resources from NASA, NSF XSEDE and Columbia University.
BWO was supported in part by the sabbatical
visitor program at the Michigan Institute for Research in Astrophysics
(MIRA) at the University of Michigan in Ann Arbor, and gratefully
acknowledges their hospitality.  Computations and associated analysis
described in this work were performed using the publicly available
Enzo code and the yt toolkit, which are the products of collaborative
efforts of many independent scientists from numerous institutions
around the world. Their commitment to open science has helped make
this work possible.


\begin{thebibliography}{}
\expandafter\ifx\csname natexlab\endcsname\relax\def\natexlab#1{#1}\fi

\bibitem[{{Baldi} {et~al.}(2009){Baldi}, {Forman}, {Jones}, {Kraft}, {Nulsen},
  {Churazov}, {David}, \& {Giacintucci}}]{Baldi2009}
{Baldi}, A., {Forman}, W., {Jones}, C., {et~al.} 2009, \apj, 707, 1034

\bibitem[{{Bertschinger} \& {Meiksin}(1986)}]{Bertschinger1986}
{Bertschinger}, E., \& {Meiksin}, A. 1986, \apjl, 306, L1

\bibitem[{{Best} {et~al.}(2007){Best}, {von der Linden}, {Kauffmann},
  {Heckman}, \& {Kaiser}}]{Best2007}
{Best}, P.~N., {von der Linden}, A., {Kauffmann}, G., {Heckman}, T.~M., \&
  {Kaiser}, C.~R. 2007, \mnras, 379, 894

\bibitem[{{B{\^i}rzan} {et~al.}(2012){B{\^i}rzan}, {Rafferty}, {Nulsen},
  {McNamara}, {R{\"o}ttgering}, {Wise}, \& {Mittal}}]{Birzan2012}
{B{\^i}rzan}, L., {Rafferty}, D.~A., {Nulsen}, P.~E.~J., {et~al.} 2012, \mnras,
  427, 3468

\bibitem[{{Blanton} {et~al.}(2011){Blanton}, {Randall}, {Clarke}, {Sarazin},
  {McNamara}, {Douglass}, \& {McDonald}}]{Blanton2011}
{Blanton}, E.~L., {Randall}, S.~W., {Clarke}, T.~E., {et~al.} 2011, \apj, 737,
  99

\bibitem[{{Bregman} \& {David}(1989)}]{Bregman1989}
{Bregman}, J.~N., \& {David}, L.~P. 1989, \apj, 341, 49

\bibitem[{{Brighenti} \& {Mathews}(2006)}]{Brighenti2006}
{Brighenti}, F., \& {Mathews}, W.~G. 2006, \apj, 643, 120

\bibitem[{{Brighenti} {et~al.}(2015){Brighenti}, {Mathews}, \&
  {Temi}}]{Brighenti2015}
{Brighenti}, F., {Mathews}, W.~G., \& {Temi}, P. 2015, ArXiv e-prints,
  arXiv:1501.07647

\bibitem[{{Bryan} {et~al.}(2014){Bryan}, {Norman}, {O'Shea}, {Abel}, {Wise},
  {Turk}, {Reynolds}, {Collins}, {Wang}, {Skillman}, {Smith}, {Harkness},
  {Bordner}, {Kim}, {Kuhlen}, {Xu}, {Goldbaum}, {Hummels}, {Kritsuk}, {Tasker},
  {Skory}, {Simpson}, {Hahn}, {Oishi}, {So}, {Zhao}, {Cen}, {Li}, \& {Enzo
  Collaboration}}]{Enzo}
{Bryan}, G.~L., {Norman}, M.~L., {O'Shea}, B.~W., {et~al.} 2014, \apjs, 211, 19

\bibitem[{{Cardiel} {et~al.}(1995){Cardiel}, {Gorgas}, \&
  {Aragon-Salamanca}}]{Cardiel1995}
{Cardiel}, N., {Gorgas}, J., \& {Aragon-Salamanca}, A. 1995, \mnras, 277, 502

\bibitem[{{Cardiel} {et~al.}(1998){Cardiel}, {Gorgas}, \&
  {Aragon-Salamanca}}]{Cardiel1998}
---. 1998, \mnras, 298, 977

\bibitem[{{Cattaneo} \& {Teyssier}(2007)}]{Cattaneo2007}
{Cattaneo}, A., \& {Teyssier}, R. 2007, \mnras, 376, 1547

\bibitem[{{Cavagnolo} {et~al.}(2008){Cavagnolo}, {Donahue}, {Voit}, \&
  {Sun}}]{Cavagnolo2008}
{Cavagnolo}, K.~W., {Donahue}, M., {Voit}, G.~M., \& {Sun}, M. 2008, \apjl,
  683, L107

\bibitem[{{Cavagnolo} {et~al.}(2010){Cavagnolo}, {McNamara}, {Nulsen},
  {Carilli}, {Jones}, \& {B{\^i}rzan}}]{Cavagnolo2010}
{Cavagnolo}, K.~W., {McNamara}, B.~R., {Nulsen}, P.~E.~J., {et~al.} 2010, \apj,
  720, 1066

\bibitem[{{Cen} \& {Ostriker}(1992)}]{CenOstriker}
{Cen}, R., \& {Ostriker}, J.~P. 1992, \apjl, 399, L113

\bibitem[{{Churazov} {et~al.}(2004){Churazov}, {Forman}, {Jones}, {Sunyaev}, \&
  {B{\"o}hringer}}]{Churazov}
{Churazov}, E., {Forman}, W., {Jones}, C., {Sunyaev}, R., \& {B{\"o}hringer},
  H. 2004, \mnras, 347, 29

\bibitem[{{Conroy} \& {Ostriker}(2008)}]{Conroy}
{Conroy}, C., \& {Ostriker}, J.~P. 2008, \apj, 681, 151

\bibitem[{{Crawford} {et~al.}(1999){Crawford}, {Allen}, {Ebeling}, {Edge}, \&
  {Fabian}}]{Crawford1999}
{Crawford}, C.~S., {Allen}, S.~W., {Ebeling}, H., {Edge}, A.~C., \& {Fabian},
  A.~C. 1999, \mnras, 306, 857

\bibitem[{{Dalla Vecchia} \& {Schaye}(2008)}]{Dalla2008}
{Dalla Vecchia}, C., \& {Schaye}, J. 2008, \mnras, 387, 1431

\bibitem[{{Dubois} {et~al.}(2010){Dubois}, {Devriendt}, {Slyz}, \&
  {Teyssier}}]{Dubois2010}
{Dubois}, Y., {Devriendt}, J., {Slyz}, A., \& {Teyssier}, R. 2010, \mnras, 409,
  985

\bibitem[{{Dunn} \& {Fabian}(2006)}]{Dunn2006}
{Dunn}, R.~J.~H., \& {Fabian}, A.~C. 2006, \mnras, 373, 959

\bibitem[{{Edge}(2001)}]{Edge2001}
{Edge}, A.~C. 2001, \mnras, 328, 762

\bibitem[{{Edwards} {et~al.}(2007){Edwards}, {Hudson}, {Balogh}, \&
  {Smith}}]{Edwards2007}
{Edwards}, L.~O.~V., {Hudson}, M.~J., {Balogh}, M.~L., \& {Smith}, R.~J. 2007,
  \mnras, 379, 100

\bibitem[{{Edwards} {et~al.}(2009){Edwards}, {Robert}, {Moll{\'a}}, \&
  {McGee}}]{Edwards2009}
{Edwards}, L.~O.~V., {Robert}, C., {Moll{\'a}}, M., \& {McGee}, S.~L. 2009,
  \mnras, 396, 1953

\bibitem[{{Fabian}(1994)}]{Fabian1994}
{Fabian}, A.~C. 1994, \araa, 32, 277

\bibitem[{{Fabian} {et~al.}(2008){Fabian}, {Johnstone}, {Sanders}, {Conselice},
  {Crawford}, {Gallagher}, \& {Zweibel}}]{Fabian2008}
{Fabian}, A.~C., {Johnstone}, R.~M., {Sanders}, J.~S., {et~al.} 2008, \nat,
  454, 968

\bibitem[{{Fabian} \& {Nulsen}(1977)}]{Fabian1977}
{Fabian}, A.~C., \& {Nulsen}, P.~E.~J. 1977, \mnras, 180, 479

\bibitem[{{Fabian} {et~al.}(2003){Fabian}, {Sanders}, {Crawford}, {Conselice},
  {Gallagher}, \& {Wyse}}]{Fabian2003}
{Fabian}, A.~C., {Sanders}, J.~S., {Crawford}, C.~S., {et~al.} 2003, \mnras,
  344, L48

\bibitem[{{Fabian} {et~al.}(2006){Fabian}, {Sanders}, {Taylor}, {Allen},
  {Crawford}, {Johnstone}, \& {Iwasawa}}]{Fabian2006}
{Fabian}, A.~C., {Sanders}, J.~S., {Taylor}, G.~B., {et~al.} 2006, \mnras, 366,
  417

\bibitem[{{Ferland} {et~al.}(2013){Ferland}, {Porter}, {van Hoof}, {Williams},
  {Abel}, {Lykins}, {Shaw}, {Henney}, \& {Stancil}}]{Cloudy}
{Ferland}, G.~J., {Porter}, R.~L., {van Hoof}, P.~A.~M., {et~al.} 2013, RMxAA,
  49, 137

\bibitem[{{Gao} \& {Solomon}(2004)}]{GaoYu2004}
{Gao}, Y., \& {Solomon}, P.~M. 2004, \apj, 606, 271

\bibitem[{{Gaspari} {et~al.}(2011){Gaspari}, {Melioli}, {Brighenti}, \&
  {D'Ercole}}]{Gaspari2011}
{Gaspari}, M., {Melioli}, C., {Brighenti}, F., \& {D'Ercole}, A. 2011, \mnras,
  411, 349

\bibitem[{{Gaspari} {et~al.}(2013){Gaspari}, {Ruszkowski}, \&
  {Oh}}]{Gaspari2013}
{Gaspari}, M., {Ruszkowski}, M., \& {Oh}, S.~P. 2013, \mnras, 432, 3401

\bibitem[{{Gaspari} {et~al.}(2012){Gaspari}, {Ruszkowski}, \&
  {Sharma}}]{Gaspari2012}
{Gaspari}, M., {Ruszkowski}, M., \& {Sharma}, P. 2012, \apj, 746, 94

\bibitem[{{Guo} \& {Oh}(2008)}]{GO2008}
{Guo}, F., \& {Oh}, S.~P. 2008, \mnras, 384, 251

\bibitem[{{Guo} \& {Oh}(2009)}]{Guo2009}
---. 2009, \mnras, 400, 1992

\bibitem[{{Guo} {et~al.}(2008){Guo}, {Oh}, \& {Ruszkowski}}]{Guo2008}
{Guo}, F., {Oh}, S.~P., \& {Ruszkowski}, M. 2008, \apj, 688, 859

\bibitem[{{Heinz} {et~al.}(2006){Heinz}, {Br{\"u}ggen}, {Young}, \&
  {Levesque}}]{Heinz2006}
{Heinz}, S., {Br{\"u}ggen}, M., {Young}, A., \& {Levesque}, E. 2006, \mnras,
  373, L65

\bibitem[{{Hickox} {et~al.}(2014){Hickox}, {Mullaney}, {Alexander}, {Chen},
  {Civano}, {Goulding}, \& {Hainline}}]{Hickox2014}
{Hickox}, R.~C., {Mullaney}, J.~R., {Alexander}, D.~M., {et~al.} 2014, \apj,
  782, 9

\bibitem[{{Hicks} \& {Mushotzky}(2005)}]{Hicks2005}
{Hicks}, A.~K., \& {Mushotzky}, R. 2005, \apjl, 635, L9

\bibitem[{{Hicks} {et~al.}(2010){Hicks}, {Mushotzky}, \& {Donahue}}]{Hicks2010}
{Hicks}, A.~K., {Mushotzky}, R., \& {Donahue}, M. 2010, \apj, 719, 1844

\bibitem[{{Hoffer} {et~al.}(2012){Hoffer}, {Donahue}, {Hicks}, \&
  {Barthelemy}}]{Hoffer2012}
{Hoffer}, A.~S., {Donahue}, M., {Hicks}, A., \& {Barthelemy}, R.~S. 2012,
  \apjs, 199, 23

\bibitem[{{Hummels} \& {Bryan}(2012)}]{Hummels}
{Hummels}, C.~B., \& {Bryan}, G.~L. 2012, \apj, 749, 140

\bibitem[{{Johnstone} {et~al.}(1987){Johnstone}, {Fabian}, \&
  {Nulsen}}]{Johnstone1987}
{Johnstone}, R.~M., {Fabian}, A.~C., \& {Nulsen}, P.~E.~J. 1987, \mnras, 224,
  75

\bibitem[{{Katz}(1992)}]{Katz1992}
{Katz}, N. 1992, \apj, 391, 502

\bibitem[{{Kennicutt}(1998)}]{Kennicutt1998}
{Kennicutt}, Jr., R.~C. 1998, \araa, 36, 189

\bibitem[{{Kim} {et~al.}(2005){Kim}, {El-Zant}, \& {Kamionkowski}}]{Kim2005}
{Kim}, W.-T., {El-Zant}, A.~A., \& {Kamionkowski}, M. 2005, \apj, 632, 157

\bibitem[{{Kritsuk} {et~al.}(2011){Kritsuk}, {Norman}, \&
  {Wagner}}]{Kritsuk2011}
{Kritsuk}, A.~G., {Norman}, M.~L., \& {Wagner}, R. 2011, \apjl, 727, L20

\bibitem[{{Lauer} {et~al.}(2005){Lauer}, {Faber}, {Gebhardt}, {Richstone},
  {Tremaine}, {Ajhar}, {Aller}, {Bender}, {Dressler}, {Filippenko}, {Green},
  {Grillmair}, {Ho}, {Kormendy}, {Magorrian}, {Pinkney}, \&
  {Siopis}}]{Lauer2005}
{Lauer}, T.~R., {Faber}, S.~M., {Gebhardt}, K., {et~al.} 2005, \aj, 129, 2138

\bibitem[{{Lee} {et~al.}(2014){Lee}, {Chang}, \& {Murray}}]{Lee2014}
{Lee}, E.~J., {Chang}, P., \& {Murray}, N. 2014, ArXiv e-prints,
  arXiv:1406.4148

\bibitem[{{Li} \& {Bryan}(2012)}]{P1}
{Li}, Y., \& {Bryan}, G.~L. 2012, \apj, 747, 26

\bibitem[{{Li} \& {Bryan}(2014{\natexlab{a}})}]{PIII}
---. 2014{\natexlab{a}}, \apj, 789, 54

\bibitem[{{Li} \& {Bryan}(2014{\natexlab{b}})}]{PII}
---. 2014{\natexlab{b}}, \apj, 789, 153

\bibitem[{{Lodato}(2007)}]{Lodato2007}
{Lodato}, G. 2007, Nuovo Cimento Rivista Serie, 30, 293

\bibitem[{{Loken} {et~al.}(2002){Loken}, {Norman}, {Nelson}, {Burns}, {Bryan},
  \& {Motl}}]{Universal_T}
{Loken}, C., {Norman}, M.~L., {Nelson}, E., {et~al.} 2002, \apj, 579, 571

\bibitem[{{Martizzi} {et~al.}(2012){Martizzi}, {Teyssier}, \&
  {Moore}}]{Martizzi2012}
{Martizzi}, D., {Teyssier}, R., \& {Moore}, B. 2012, \mnras, 420, 2859

\bibitem[{{Mathews} {et~al.}(2006){Mathews}, {Faltenbacher}, \&
  {Brighenti}}]{Mathews}
{Mathews}, W.~G., {Faltenbacher}, A., \& {Brighenti}, F. 2006, \apj, 638, 659

\bibitem[{{McCourt} {et~al.}(2012){McCourt}, {Sharma}, {Quataert}, \&
  {Parrish}}]{McCourt12}
{McCourt}, M., {Sharma}, P., {Quataert}, E., \& {Parrish}, I.~J. 2012, \mnras,
  419, 3319

\bibitem[{{McDonald} {et~al.}(2013){McDonald}, {Benson}, {Veilleux}, {Bautz},
  \& {Reichardt}}]{Phoenix}
{McDonald}, M., {Benson}, B., {Veilleux}, S., {Bautz}, M.~W., \& {Reichardt},
  C.~L. 2013, \apjl, 765, L37

\bibitem[{{McDonald} {et~al.}(2011{\natexlab{a}}){McDonald}, {Veilleux}, \&
  {Mushotzky}}]{McDonald11}
{McDonald}, M., {Veilleux}, S., \& {Mushotzky}, R. 2011{\natexlab{a}}, \apj,
  731, 33

\bibitem[{{McDonald} {et~al.}(2012){McDonald}, {Veilleux}, \&
  {Rupke}}]{McDonald12}
{McDonald}, M., {Veilleux}, S., \& {Rupke}, D.~S.~N. 2012, \apj, 746, 153

\bibitem[{{McDonald} {et~al.}(2010){McDonald}, {Veilleux}, {Rupke}, \&
  {Mushotzky}}]{McDonald10}
{McDonald}, M., {Veilleux}, S., {Rupke}, D.~S.~N., \& {Mushotzky}, R. 2010,
  \apj, 721, 1262

\bibitem[{{McDonald} {et~al.}(2011{\natexlab{b}}){McDonald}, {Veilleux},
  {Rupke}, {Mushotzky}, \& {Reynolds}}]{McDonald11b}
{McDonald}, M., {Veilleux}, S., {Rupke}, D.~S.~N., {Mushotzky}, R., \&
  {Reynolds}, C. 2011{\natexlab{b}}, \apj, 734, 95

\bibitem[{{McNamara} \& {Nulsen}(2007)}]{McNamara2007}
{McNamara}, B.~R., \& {Nulsen}, P.~E.~J. 2007, \araa, 45, 117

\bibitem[{{McNamara} \& {O'Connell}(1989)}]{McNamara1989}
{McNamara}, B.~R., \& {O'Connell}, R.~W. 1989, \aj, 98, 2018

\bibitem[{{Merritt} \& {Ferrarese}(2001)}]{Msigma}
{Merritt}, D., \& {Ferrarese}, L. 2001, \mnras, 320, L30

\bibitem[{{Mittal} {et~al.}(2009){Mittal}, {Hudson}, {Reiprich}, \&
  {Clarke}}]{Mittal2009}
{Mittal}, R., {Hudson}, D.~S., {Reiprich}, T.~H., \& {Clarke}, T. 2009, \aap,
  501, 835

\bibitem[{{Narayan} \& {Medvedev}(2001)}]{Narayan2001}
{Narayan}, R., \& {Medvedev}, M.~V. 2001, \apjl, 562, L129

\bibitem[{{Navarro} {et~al.}(1996){Navarro}, {Frenk}, \& {White}}]{NFW}
{Navarro}, J.~F., {Frenk}, C.~S., \& {White}, S.~D.~M. 1996, \apj, 462, 563

\bibitem[{{O'Dea} {et~al.}(2008){O'Dea}, {Baum}, {Privon}, {Noel-Storr},
  {Quillen}, {Zufelt}, {Park}, {Edge}, {Russell}, {Fabian}, {Donahue},
  {Sarazin}, {McNamara}, {Bregman}, \& {Egami}}]{ODea2008}
{O'Dea}, C.~P., {Baum}, S.~A., {Privon}, G., {et~al.} 2008, \apj, 681, 1035

\bibitem[{{O'Dea} {et~al.}(2010){O'Dea}, {Quillen}, {O'Dea}, {Tremblay},
  {Snios}, {Baum}, {Christiansen}, {Noel-Storr}, {Edge}, {Donahue}, \&
  {Voit}}]{ODea2010}
{O'Dea}, K.~P., {Quillen}, A.~C., {O'Dea}, C.~P., {et~al.} 2010, \apj, 719,
  1619

\bibitem[{{Omma} {et~al.}(2004){Omma}, {Binney}, {Bryan}, \& {Slyz}}]{Omma2004}
{Omma}, H., {Binney}, J., {Bryan}, G., \& {Slyz}, A. 2004, \mnras, 348, 1105

\bibitem[{{Parrish} {et~al.}(2009){Parrish}, {Quataert}, \&
  {Sharma}}]{Parrish2009}
{Parrish}, I.~J., {Quataert}, E., \& {Sharma}, P. 2009, \apj, 703, 96

\bibitem[{{Peterson} \& {Fabian}(2006)}]{Peterson2006}
{Peterson}, J.~R., \& {Fabian}, A.~C. 2006, \physrep, 427, 1

\bibitem[{{Peterson} {et~al.}(2003){Peterson}, {Kahn}, {Paerels}, {Kaastra},
  {Tamura}, {Bleeker}, {Ferrigno}, \& {Jernigan}}]{Peterson2003}
{Peterson}, J.~R., {Kahn}, S.~M., {Paerels}, F.~B.~S., {et~al.} 2003, \apj,
  590, 207

\bibitem[{{Pizzolato} \& {Soker}(2005)}]{Pizzolato2005}
{Pizzolato}, F., \& {Soker}, N. 2005, \apj, 632, 821

\bibitem[{{Postman} {et~al.}(2012){Postman}, {Coe}, {Ben{\'{\i}}tez},
  {Bradley}, {Broadhurst}, {Donahue}, {Ford}, {Graur}, {Graves}, {Jouvel},
  {Koekemoer}, {Lemze}, {Medezinski}, {Molino}, {Moustakas}, {Ogaz}, {Riess},
  {Rodney}, {Rosati}, {Umetsu}, {Zheng}, {Zitrin}, {Bartelmann}, {Bouwens},
  {Czakon}, {Golwala}, {Host}, {Infante}, {Jha}, {Jimenez-Teja}, {Kelson},
  {Lahav}, {Lazkoz}, {Maoz}, {McCully}, {Melchior}, {Meneghetti}, {Merten},
  {Moustakas}, {Nonino}, {Patel}, {Reg{\"o}s}, {Sayers}, {Seitz}, \& {Van der
  Wel}}]{Postman2012}
{Postman}, M., {Coe}, D., {Ben{\'{\i}}tez}, N., {et~al.} 2012, \apjs, 199, 25

\bibitem[{{Rafferty} {et~al.}(2006){Rafferty}, {McNamara}, {Nulsen}, \&
  {Wise}}]{Rafferty2006}
{Rafferty}, D.~A., {McNamara}, B.~R., {Nulsen}, P.~E.~J., \& {Wise}, M.~W.
  2006, \apj, 652, 216

\bibitem[{{Rawle} {et~al.}(2012){Rawle}, {Edge}, {Egami}, {Rex}, {Smith},
  {Altieri}, {Fiedler}, {Haines}, {Pereira}, {P{\'e}rez-Gonz{\'a}lez},
  {Portouw}, {Valtchanov}, {Walth}, {van der Werf}, \& {Zemcov}}]{Rawle2012}
{Rawle}, T.~D., {Edge}, A.~C., {Egami}, E., {et~al.} 2012, \apj, 747, 29

\bibitem[{{Rosner} \& {Tucker}(1989)}]{Rosner}
{Rosner}, R., \& {Tucker}, W.~H. 1989, \apj, 338, 761

\bibitem[{{Ruszkowski} \& {Begelman}(2002)}]{Ruszkowski2002}
{Ruszkowski}, M., \& {Begelman}, M.~C. 2002, \apj, 581, 223

\bibitem[{{Ruszkowski} {et~al.}(2004){Ruszkowski}, {Br{\"u}ggen}, \&
  {Begelman}}]{Ruszkowski2004}
{Ruszkowski}, M., {Br{\"u}ggen}, M., \& {Begelman}, M.~C. 2004, \apj, 611, 158

\bibitem[{{Ruszkowski} \& {Oh}(2010)}]{RO10}
{Ruszkowski}, M., \& {Oh}, S.~P. 2010, \apj, 713, 1332

\bibitem[{{Ruszkowski} \& {Oh}(2011)}]{Ruszkowski2011}
---. 2011, \mnras, 414, 1493

\bibitem[{{Salem} \& {Bryan}(2014)}]{Munier}
{Salem}, M., \& {Bryan}, G.~L. 2014, \mnras, 437, 3312

\bibitem[{{Salom{\'e}} \& {Combes}(2003)}]{Salome2003}
{Salom{\'e}}, P., \& {Combes}, F. 2003, \aap, 412, 657

\bibitem[{{Sharma} {et~al.}(2012){Sharma}, {McCourt}, {Quataert}, \&
  {Parrish}}]{Sharma2012}
{Sharma}, P., {McCourt}, M., {Quataert}, E., \& {Parrish}, I.~J. 2012, \mnras,
  420, 3174

\bibitem[{{Sijacki} \& {Springel}(2006)}]{Sijacki2006}
{Sijacki}, D., \& {Springel}, V. 2006, \mnras, 366, 397

\bibitem[{{Simpson} {et~al.}(2014){Simpson}, {Bryan}, {Hummels}, \&
  {Ostriker}}]{Christine}
{Simpson}, C.~M., {Bryan}, G.~L., {Hummels}, C., \& {Ostriker}, J.~P. 2014,
  ArXiv e-prints, arXiv:1410.3822

\bibitem[{{Singh} \& {Sharma}(2015)}]{Sharma2015}
{Singh}, A., \& {Sharma}, P. 2015, \mnras, 446, 1895

\bibitem[{{Skory} {et~al.}(2013){Skory}, {Hallman}, {Burns}, {Skillman},
  {O'Shea}, \& {Smith}}]{Skory}
{Skory}, S., {Hallman}, E., {Burns}, J.~O., {et~al.} 2013, \apj, 763, 38

\bibitem[{{Smith} {et~al.}(2013){Smith}, {O'Shea}, {Voit}, {Ventimiglia}, \&
  {Skillman}}]{Smith2013}
{Smith}, B., {O'Shea}, B.~W., {Voit}, G.~M., {Ventimiglia}, D., \& {Skillman},
  S.~W. 2013, \apj, 778, 152

\bibitem[{{Spitzer}(1956)}]{Spitzer}
{Spitzer}, Jr., L. 1956, \apj, 124, 20

\bibitem[{{Stone} \& {Norman}(1992)}]{Zeus}
{Stone}, J.~M., \& {Norman}, M.~L. 1992, \apjs, 80, 753

\bibitem[{{Tasker} \& {Bryan}(2006)}]{Tasker2006}
{Tasker}, E.~J., \& {Bryan}, G.~L. 2006, \apj, 641, 878

\bibitem[{{Thacker} {et~al.}(2014){Thacker}, {MacMackin}, {Wurster}, \&
  {Hobbs}}]{Thacker2014}
{Thacker}, R.~J., {MacMackin}, C., {Wurster}, J., \& {Hobbs}, A. 2014, \mnras,
  443, 1125

\bibitem[{{Truelove} {et~al.}(1997){Truelove}, {Klein}, {McKee}, {Holliman},
  {Howell}, \& {Greenough}}]{Truelove}
{Truelove}, J.~K., {Klein}, R.~I., {McKee}, C.~F., {et~al.} 1997, \apjl, 489,
  L179

\bibitem[{{Turk} {et~al.}(2011){Turk}, {Smith}, {Oishi}, {Skory}, {Skillman},
  {Abel}, \& {Norman}}]{yt}
{Turk}, M.~J., {Smith}, B.~D., {Oishi}, J.~S., {et~al.} 2011, \apjs, 192, 9

\bibitem[{{Urban} {et~al.}(2014){Urban}, {Simionescu}, {Werner}, {Allen},
  {Ehlert}, {Zhuravleva}, {Morris}, {Fabian}, {Mantz}, {Nulsen}, {Sanders}, \&
  {Takei}}]{Urban2014}
{Urban}, O., {Simionescu}, A., {Werner}, N., {et~al.} 2014, \mnras, 437, 3939

\bibitem[{{Van Loo} {et~al.}(2014){Van Loo}, {Tan}, \& {Falle}}]{Loo}
{Van Loo}, S., {Tan}, J.~C., \& {Falle}, S.~A.~E.~G. 2014, ArXiv e-prints,
  arXiv:1411.7548

\bibitem[{{Vernaleo} \& {Reynolds}(2006)}]{Reynolds06}
{Vernaleo}, J.~C., \& {Reynolds}, C.~S. 2006, \apj, 645, 83

\bibitem[{{Voigt} {et~al.}(2002){Voigt}, {Schmidt}, {Fabian}, {Allen}, \&
  {Johnstone}}]{Voigt}
{Voigt}, L.~M., {Schmidt}, R.~W., {Fabian}, A.~C., {Allen}, S.~W., \&
  {Johnstone}, R.~M. 2002, \mnras, 335, L7

\bibitem[{{Voit}(2011)}]{Voit2011}
{Voit}, G.~M. 2011, \apj, 740, 28

\bibitem[{{Voit} \& {Donahue}(2014)}]{MarkMegan2014}
{Voit}, G.~M., \& {Donahue}, M. 2014, ArXiv e-prints, arXiv:1409.1601

\bibitem[{{Voit} {et~al.}(2014){Voit}, {Donahue}, {Bryan}, \&
  {McDonald}}]{Mark2014Nature1}
{Voit}, G.~M., {Donahue}, M., {Bryan}, G.~L., \& {McDonald}, M. 2014, ArXiv
  e-prints, arXiv:1409.1598

\bibitem[{{Wilman} {et~al.}(2005){Wilman}, {Edge}, \& {Johnstone}}]{BHmass}
{Wilman}, R.~J., {Edge}, A.~C., \& {Johnstone}, R.~M. 2005, \mnras, 359, 755

\bibitem[{{Wise} {et~al.}(2007){Wise}, {McNamara}, {Nulsen}, {Houck}, \&
  {David}}]{Wise2007}
{Wise}, M.~W., {McNamara}, B.~R., {Nulsen}, P.~E.~J., {Houck}, J.~C., \&
  {David}, L.~P. 2007, \apj, 659, 1153

\bibitem[{{Zakamska} \& {Narayan}(2003)}]{Narayan2003}
{Zakamska}, N.~L., \& {Narayan}, R. 2003, \apj, 582, 162

\end{thebibliography}
\end{document}